\documentclass[aps,prd,showpacs,twocolumn,superscriptaddress,nofootinbib,preprintnumbers]{revtex4-1} 

\makeatletter
\def\p@subsection{}
\makeatother

\usepackage{color,graphicx,amsmath,amssymb,lipsum,bm}
\usepackage[colorlinks=true,citecolor=blue,linkcolor=blue,urlcolor=blue, backref=false,pdfborder={0 0 0}]{hyperref}

\usepackage[normalem]{ulem}
\usepackage[utf8]{inputenc} 
\usepackage{mathtools}

\usepackage{float}

\usepackage{ulem}
\usepackage{diagbox}

\usepackage[dvipsnames]{xcolor}
\usepackage{mathrsfs}

\newcommand{\be}{\begin{equation}}
\newcommand{\ee}{\end{equation}}
\newcommand{\beqa}{\begin{eqnarray}}
\newcommand{\eeqa}{\end{eqnarray}}

\newcommand\p{{\bm p}}
\renewcommand\k{{\bm k}}

\newcommand\GG{\Gamma_3}
\newcommand\G{\mathcal{G}_2}

\newcommand{\bseq}{\begin{subequations}}
\newcommand{\eseq}{\end{subequations}}

\renewcommand{\ln}{\mathop{\rm ln}\nolimits}

\def\gsim{\raise0.3ex\hbox{$\;>$\kern-0.75em\raise-1.1ex\hbox{$\sim\;$}}}
\def\lsim{\raise0.3ex\hbox{$\;<$\kern-0.75em\raise-1.1ex\hbox{$\sim\;$}}}

\def\beqn#1{\begin{equation}\label{#1}}
\def\eeqn{\end{equation}}

\def\beqa#1{\begin{eqnarray}\label{#1}}
\def\eeqa{\end{eqnarray}}

\def\Z2{$\mathcal{Z_2}$}

\newcommand {\ignore}[1]{}

\renewcommand{\arraystretch}{1.3} 
 
\usepackage{multirow}

\begin{document}

\preprint{MIT-CTP/5881}

\title{Constraining Dynamical Dark Energy from Galaxy Clustering \\
with 
Simulation-Based Priors}

\author{Shu-Fan Chen} 
\email{shufan\_chen@g.harvard.edu
 } 
\affiliation{Department of Physics, Harvard University, Cambridge, MA 02138, USA}

\author{Mikhail M. Ivanov }
\email{ivanov99@mit.edu}
\affiliation{Center for Theoretical Physics -- a Leinweber Institute, Massachusetts Institute of Technology, 
Cambridge, MA 02139, USA} 
 \affiliation{The NSF AI Institute for Artificial Intelligence and Fundamental Interactions, Cambridge, MA 02139, USA}

\begin{abstract} 
The effective-field theory based full-shape analysis
with simulation-based priors (EFT-SBP) is the novel 
analysis of galaxy clustering data that allows
one 
to combine merits of perturbation theory 
and simulation-based modeling in a unified framework. In this paper we use EFT-SBP with the galaxy clustering power spectrum and bispectrum data from 
BOSS in order to test the recent preference for dynamical dark energy reported by the DESI collaboration. 
While dynamical dark energy is preferred by the combination of DESI baryon acoustic oscillation, \textit{Planck} Cosmic Microwave Background, 
and Pantheon+ supernovae data, we show that this preference disappears once these data sets are combined with 
the usual BOSS EFT galaxy power spectrum and bispectrum 
likelihood. The use of the simulation-based priors in this analysis further
weakens the case for dynamical dark energy by additionally 
shrinking the parameter
posterior around the cosmological 
constant region. 
Specifically, the figure of merit of the dynamical dark energy constraints 
from the combined data set 
improves by $\approx 20\%$ over the usual EFT-full-shape analysis with the conservative priors.
These results are made possible with a novel 
modeling approach to the EFT 
prior distribution 
with the Gaussian mixture models,
which allows us to both accurately 
capture the EFT priors and retain 
the ability to analytically marginalize
the likelihood 
over most of the EFT nuisance parameters. 
Our results 
challenge the dynamical dark energy interpretation
of the DESI data and enable future 
EFT-SBP analyses of BOSS and DESI in the context of non-minimal cosmological models. 
\end{abstract}

\maketitle

\section{Introduction}

The standard cosmological 
model, inflationary $\Lambda$CDM, 
provides a first approximation
to the basic properties and the evolution of our 
Universe. This model assumes 
the presence of three new physics entities: the exponential 
primordial accelerated 
expansion of the Universe (dubbed cosmic inflation), 
the cosmological constant to 
explain the current accelerated
expansion of the Universe, 
and dark matter in order to account
for the observed cosmic structure 
on both large and small scales. 
The precise nature of these
phenomena remains the subject 
of intense observational and 
theoretical efforts. 
In addition to that, there are several observational tensions
suggesting the breakdown 
of $\Lambda$CDM, e.g. the 
evidence for dynamical
dark energy recently reported
by the Dark Energy Survey 
Instrument (DESI) collaboration~\cite{DESI:2024mwx,DESI:2024mwx,DESI:2024uvr,DESI:2025zgx,DESI:2025fii}. 
The fate of $\Lambda$CDM 
will depend on the outcome 
of ongoing and future 
large-scale structure 
galaxy surveys,
such as DESI, Euclid~\cite{Laureijs:2011gra}, LSST~\cite{LSST:2008ijt}, and Roman Space Telescope~\cite{Akeson:2019biv}.

The cosmological interpretation of data from 
galaxy surveys is obscured by 
effects of non-linear structure formation. There are two leading 
methods to model these effects. 
The first one is to simulate the 
formation of galaxies numerically by consistently solving 
a set of equations governing the 
gravitational collapse of dark matter and baryons, supplemented with a closure prescription. This method is exemplified by large-scale hydrodynamical simulations such 
EAGLE, IllustrisTNG
and MillenniumTNG simulations~\cite{McAlpine:2015tma,Springel:2017tpz,Hernandez-Aguayo:2022xcl}.
Impressive as they are, a large computational cost of these simulations prevents them 
from being used for the inference 
of cosmological parameters from data. An ersatz version of
full hydrodynamical simulations 
is provided by the halo-occupation 
distribution (HOD) 
framework~\cite{Berlind:2001xk,Kravtsov:2003sg,Zheng:2004id,Hearin:2015jnf,Wechsler:2018pic}, where
one self-consistently simulates 
only the clustering of dark matter via the N-body method, 
and the galaxies are ``painted''
on top of dark matter halos 
according so a certain model
probability distribution whose
form is either tuned to reproduce
the observational data or 
the full simulation results. 
The success of the HOD-based
analyses methods 
is exemplified by 
the Beyond-2pt Community Data Challenge~\cite{Beyond-2pt:2024mqz},
and by re-analyses of galaxy clustering data from the Baryon acoustic Oscillation Spectroscopic Survey (BOSS~\cite{BOSS:2016wmc})~\cite{Kobayashi:2021oud,Cuesta-Lazaro:2023gbv,Valogiannis:2023mxf,Hahn:2023kky,Hou:2024blc}. 

The second method to model 
non-linear structure formation is to use perturbation theory, 
whose recent incarnation 
is known as the effective field
theory of large-scale structure (EFT)~\cite{Baumann:2010tm,Carrasco:2012cv,Ivanov:2022mrd}.
In EFT, the correlation between
the distribution of galaxies 
and the underlying distribution
of dark matter 
on large scales is described by 
an expansion based on symmetries and dimensional analysis, called the bias expansion~\cite{Desjacques:2016bnm}.
The details of galaxy formation
are parameterized by 
the so-called EFT parameters,
which are generalizations of classic
perturbative bias parameters. 
In cosmological analyses, 
these parameters have to be marginalized over~\cite{Ivanov:2019pdj,DAmico:2019fhj,Chen:2021wdi,Philcox:2021kcw,Ivanov:2023qzb,Chen:2024vuf}. 
While EFT provides a first
principle agnostic description to clustering of galaxies on large-scales, it misses the information
encoded in the small-scale clustering properties. The latter, however, can be extracted from the simulations. 
The simplest approach to galaxy clustering analysis that combines
the merits of EFT on large scales
and simulations on small scales
is the EFT-based full-shape 
analysis with simulation-based
priors (EFT-SBP)~\cite{Ivanov:2024hgq,Ivanov:2024xgb,Cabass:2024wob,Ivanov:2024dgv,Ivanov:2025qie,Sullivan:2025eei} (see also~\cite{Sullivan:2021sof,Obuljen:2022cjo,Modi:2023drt,Akitsu:2024lyt,Zhang:2024thl,DESI:2025wzd,Zhang:2025sfk}). In this approach, the 
EFT-based analysis is augmented 
by non-perturbative information
on galaxy formation from small scales implemented as priors on EFT parameters. 
So far, these priors are extracted from
large sets of HOD-based catalogs,
although the original method~\cite{Ivanov:2024hgq,Ivanov:2024xgb}
can be applied to the full 
hydrodynamical simulations as well~\cite{Ivanov:2024dgv}. 
In the original approach, the
EFT parameters 
are calibrated at the field-level, which allows for their precision measurements
by means of the sample variance
cancellation~\cite{Schmittfull:2018yuk,Schmittfull:2020trd}.\footnote{See also refs.~\cite{Obuljen:2022cjo,Foreman:2024kzw} for the application to HI maps, 
and~\cite{Schmittfull:2014tca,Lazeyras:2017hxw,
Abidi:2018eyd,
Schmidt:2018bkr,
Elsner:2019rql,
Cabass:2019lqx,
Modi:2019qbt,
Schmidt:2020tao,
Schmidt:2020viy,
Lazeyras:2021dar,
Stadler:2023hea,
Nguyen:2024yth} for additional 
references on field-level EFT.} (Recently, there also have been efforts along the same lines 
with EFT parameters calibrated 
at the level of the galaxy power spectrum without the sample variance
cancellation~\cite{Zhang:2024thl,DESI:2025wzd}.) 

The distribution of EFT parameters
extracted from the simulations
is highly non-Gaussian. So far, it has been modeled by means 
of normalizing flows~\cite{Ivanov:2024hgq,Ivanov:2024xgb}, 
a machine learning tool
capable of fitting non-Gaussian distributions by numerically 
mapping them onto the Gaussian ones. This method has a disadvantage that all EFT parameters
of galaxies have to be sampled explicitly in the analysis, 
which makes the analysis time consuming. Indeed, the standard EFT full-shape analysis with the conservative Gaussian 
priors on EFT parameters
can be much faster because 
the relevant likelihoods depend on the EFT parameters quadratically, and hence can be analytically 
marginalized over them. 
The need for an explicit sampling
makes it hard to run difficult analyses, e.g. to combine EFT-SBP 
with data from the CMB, and 
supernovae in the context
of the extended cosmological models.
The latter is particularly 
important if one were to 
apply EFT-SBP to analyses of
dynamical dark energy 
in the context of recent
reports by the DESI collaboration.
However, there is a significant demand for such an analysis 
given that the SBP calibrated at the field level strongly 
enhance the constraints
on the parameters of the $\Lambda$CDM model. Given these
results, 
and the potential of the EFT-full-shape data to deliver precise constraints on dynamical dark energy
even with the conservative priors~\cite{Chudaykin:2020ghx,Chen:2024vuf}, 
it is natural to 
expect significant improvements
on the dark energy constraints 
from the EFT-SBP. 

In this work, we resolve the challenge of the computational implementation of simulation-based
priors by developing a novel modeling 
approach to EFT parameter distributions based on the Gaussian 
mixture model (GMM). GMM is an alternative way to model a multi-dimensional non-Gaussian
distribution as a sum of 
Gaussian distributions. In contrast to the normalizing flows, it allows for an implementation of the simulation-based priors in a quasi-Gaussian form, thus enabling analytic marginalization~\cite{Philcox:2020zyp,Philcox:2021kcw}
and hence a significant reduction
in the analysis time for difficult
runs.

As an application of our method, 
we analyze a combination of 
large-scale structure, CMB,
and supernovae data 
in the context of dynamical
dark energy.  
We test the consistency
and robustness of the dynamical dark energy preference in the DESI's
Baryon Acoustic 
Oscillation (BAO)
and supernovae
data
w.r.t. addition of the clustering data 
from the BOSS survey 
enhanced with the SBPs. 
Our analysis has intriguing phenomenological consequences. We show that the addition of the 
EFT full-shape likelihood with the 
simulation-based priors 
reduces the evidence for dynamical 
dark energy otherwise favored by the 
combination of the DESI's DR2 BAO~\cite{DESI:2025zgx}, Planck CMB~\cite{Planck:2018vyg}, and Pantheon+ supernovae (SNe) data~\cite{Scolnic:2021amr,Brout:2022vxf}. 
This suggests that the field-level 
simulation-based priors 
implemented via GMM 
is a powerful tool
capable of significantly improving
dark energy constraints from galaxy clustering analyses. 

Our paper is structured as follows. 
In Section~\ref{sec:method} we review the 
EFT model for galaxy clustering, 
introduce the EFT 
parameters and discuss their measurements
from the halo occupation distribution
mock 
catalogs. There we discuss in detail 
the analytic marginalization procedure
for a Gaussian prior and a prior represented
by a Gaussian Mixture Model. 
In Section~\ref{sec:datasets} 
we discuss the datasets used in our analysis. 
Section~\ref{sec:res}
presents the main phenomenological results
of our work in terms of new constraints 
on the dynamical dark energy parameters.
Finally, we draw conclusions in 
Section~\ref{sec:disc}.
Some additional validation material 
is presented in the Appendix.

\section{The distribution of EFT parameters as a Gaussian mixture}\label{sec:method}

Incorporating the simulation-based prior for nuisance parameters within the EFT framework presents a significant challenge due to the sheer number of such parameters. For instance, even at the level of the power spectrum alone, there are eleven EFT nuisance parameters that must be marginalized over to constrain cosmological parameters. Under a conservative Gaussian prior, these can typically be marginalized analytically using Gaussian integrals. However, when the prior is derived from N-body + HOD simulations, the resulting distribution is highly non-Gaussian, and normalizing flows are commonly employed to approximate its complex structure.

While normalizing flows offer high fidelity in approximating the true log-likelihood, they come with notable limitations:
(1) Fast likelihood evaluation typically requires GPU acceleration because normalizing flow consists of neural network;
(2) The model can become unstable near the boundaries of the distribution if sample coverage is insufficient, leading to overfitting;
(3) All parameters must be sampled explicitly, which significantly slows down posterior inference.

To address these issues, we instead adopt Gaussian Mixture Models (GMMs)—representing the prior as a weighted sum of multiple Gaussian distributions. GMMs offer a flexible yet tractable approximation to the complex simulation-based prior. Crucially, since GMMs are built from Gaussian components, they retain the desirable property of allowing for analytic marginalization of parameters that enter linearly. In this section, we will
recap EFT parameters, and then continue by reviewing the analytic marginalization procedure for a single multivariate Gaussian, and then extend it to the case of Gaussian mixture models.

\subsection{Recap of EFT modeling and parameters}

We begin with the Eulerian bias expansion for the galaxy overdensity field in the Effective Field Theory (EFT) framework~\cite{Ivanov:2019pdj,Ivanov:2019hqk,Chudaykin:2020hbf,Ivanov:2021kcd}:
\be 
\label{eq:dg}
\begin{split}
    \delta_{g,\rm E}^{\rm EFT}(\k) = &b_1\delta + \frac{b_2}{2}\delta^2 + b_{\mathcal{G}_2}\mathcal{G}_2 + b_{\Gamma_3}\Gamma_3 - b_{\nabla^2\delta}\nabla^2\delta \\
    & + \epsilon\,,
    \end{split}
\ee
where $\delta$ is the non-linear matter density field, $\mathcal{G}_2$ is the tidal operator, and $\Gamma_3$ is the Galileon-type tidal operator. These higher-order terms are defined as
\be 
\begin{split}
  &   \mathcal{S}_2(\k) = \int_\p F_{\mathcal{S}_2}(\p,\k-\p)\,\delta(\p)\,\delta(\k-\p)\,, \\
    & \mathcal{S}_2 \in \{\mathcal{G}_2,\,\delta^2/2\}\,, \\
   & F_{\mathcal{G}_2}(\k_1,\k_2) = \frac{(\k_1\cdot\k_2)^2}{k_1^2k_2^2} - 1\,, \quad  F_{\delta^2/2}(\k_1,\k_2)=\frac{1}{2}\,, \\
   & \Gamma_3(\k) = \left(\prod_{n=1}^3\int_{\k_n} \delta(\k_n)\right)
  (2\pi)^3\delta_D^{(3)}(\k-\k_{123})
    F_{\Gamma_3}(\k_1,\k_2,\k_3)\,, \\
   & F_{\Gamma_3}(\k_1,\k_2,\k_3) = \frac{4}{7}\left(1 - \frac{(\k_1\cdot\k_2)^2}{k_1^2k_2^2}\right)\left(\frac{[\k_{12}\cdot\k_3]^2}{|\k_{12}|^2\,k_3^2} - 1\right)\,.
    \end{split}
\ee
where $\k_{1...n}\equiv \k_1+...+\k_n$.
$\epsilon$ above is the stochastic 
field that produces the shot noise
contribution on large scales,
while $\nabla^2\delta$
is the higher derivative bias parameter. 
The non-linear field $\delta$
is subject to a perturbative expansion
over the linear matter density field
at the one loop order:
\be 
\delta=\sum_{n=1}^3\left(\prod_{i=1}^n
\int_{\k_i}\delta_1(\k_i)\right)
(2\pi)^3\delta_D^{(3)}(\k-\k_{1...n})F_n(\k_1,...,\k_n)\,,
\ee 
where $F_n$ is the matter
density kernel 
in  standard perturbation theory~\cite{Bernardeau:2001qr}.
$\delta_1$ is assumed to be a Gaussian
field characterized by its power
spectrum
\be 
\langle 
\delta_1(\k)\delta_1(\k')\rangle=(2\pi)^3\delta_D^{(3)}(\k+\k')D^2(z)P_{11}(k)\,,
\ee 
where $D$ is the linear theory
growth factor. 

Eq.~\eqref{eq:dg}
is subject to redshift space
mapping~\cite{Bernardeau:2001qr}, which produces the dependence
on the line of sight, so that the galaxy
power spectrum will become a function
of the wavevector module $k$
and its line of sight projection $\mu\equiv (\k\cdot \hat{\bf z})/k$.
In that case the power spectrum
is decomposed 
into the multipole moments,
\be 
P(k,\mu,z) = \sum_{\ell=0,2,4}P_\ell(k,z)L_\ell(\mu)\,,
\ee 
where $L_\ell$
is the Legendre polynomial of order $\ell$, 
and we only focus on the 
first three 
moments $(0,2,4)$,
which dominate the signal. 
The renormalization of the redshift-space mapping produces extra 
higher-derivative counterterms.
Their contributions 
to the power spectrum multipoles can 
written as~\cite{Chudaykin:2020aoj}
\be 
P^{\rm ctr.}_\ell(k)=- c_{s,\ell} k^2 \frac{2\ell+1}{2}\int_{-1}^1d\mu (f\mu^2)^\frac{\ell}{2} L_\ell(\mu)P_{11}(k)\,,
\ee 
where $f\equiv d\log D/d\log a$,
and $a$ is the metric scale factor.
Matching the 
convention used 
by~\cite{Ivanov:2024hgq} is realized by
$c_{s,\ell}=c_\ell$ there. 
In addition to that, 
we also consider a higher-derivative
improvement of the redshift-space
power spectrum~\cite{Ivanov:2019pdj,Taule:2023izt} 
\be 
P_{\nabla^4_z\delta}(k,\mu)=-b_4 k^4 \mu^4 f^4 (b_1+f\mu^2)^2 P_{11}(k)\,,
\ee
which captures the deterministic 
part of non-linear redshift space distortions 
(fingers-of-God)~\cite{Jackson:2008yv}.
Note that in some literature
$b_4$
is denoted as $\tilde{c}$.

For the stochastic terms that are independently distributed relative to the above the density field, the EFT
prediction is~\cite{Perko:2016puo,Chudaykin:2020aoj}
\be 
    P_{\rm stoch}(k,\mu) = \frac{1+P_{\rm shot} }{\bar{n}}+ (a_0 + a_2\mu^2)\left(\frac{k}{k_S}\right)^2\,,
\ee
where $\bar{n}$ is the number density of galaxies and $k_S=0.45$ $h/$Mpc
is a normalization scale. Matching the 
convention used 
by~\cite{Ivanov:2024hgq} is realized by
\be 
P_{\rm shot}=\alpha_0\,,a_0=\alpha_1\,,a_2=\alpha_2\,.
\ee 
These 
parameters have been 
measured
from large catalogs of 
the HOD galaxies
in~\cite{Ivanov:2024xgb}.
These measurements
constitute the 
simulations-based
priors which we will
use this work. 

All in all, the one-loop EFT model depends on 
11 parameters: the bias parameters 
$\{b_1,b_2,b_{\G},b_{\GG}\}$,
the counterterms 
$\{c_{s,0},c_{s,2},c_{s,4},b_4\}$,
and the stochasticity parameters 
$\{P_{\rm shot}, a_0,a_2\}$.
Eight of these parameters, 
$\{b_{\GG},c_{s,0},c_{s,2},c_{s,4},b_4,P_{\rm shot}, a_0,a_2\}$,
enter the model linearly
and hence the likelihood quadratically,
so that they can be 
analytically marginalized over
if the prior is Gaussian. 
Let us discuss this 
case now. 

\subsection{Analytical Marginalization of Gaussian Likelihood}
Here, we will derive the analytical marginalization of a Gaussian likelihood over nuisance parameters that enter linearly and have Gaussian priors, allowing for the case when there are correlations between linearly ($\theta_l$) and non-linearly ($\theta_{n}$) entering parameters. If the model is
\be 
    d = m(\theta_{n}) + X(\theta_{n}) \theta_l
\ee
and if we assume that the joint prior is Gaussian
\be 
    p(\theta) \sim \mathcal{N}(\mu, \Sigma), \quad \theta = \begin{pmatrix} \theta_l \\ \theta_n \end{pmatrix}, \quad \mu = \begin{pmatrix} \mu_l \\ \mu_n \end{pmatrix},
\ee
with prior covariance
\begin{align}
    \Sigma = \begin{pmatrix} \Sigma_{ll} & \Sigma_{ln} \\ \Sigma_{ln}^T & \Sigma_{nn} \end{pmatrix}\,,
\end{align}
then the conditional prior for the linear parameters given the nonlinear ones can be written as
\begin{align}
    \mu_{l | n} &= \mu_l + \Sigma_{ln} \Sigma_{nn}^{-1} (\theta_n - \mu_n) \\
    \Sigma_{l | n} &= \Sigma_{ll} - \Sigma_{ln} \Sigma_{nn}^{-1} \Sigma_{ln}^T
\end{align}
With this setup, we will start the analytic marginalization for the likelihood function. The full likelihood function, including the prior is
\begin{widetext}
\begin{align*}
    \mathcal{L}(\theta) \propto \exp\left[-\frac{1}{2}(d - m(\theta_n) - X(\theta_n) \theta_l)^T C^{-1} (d - m(\theta_n) - X(\theta_n) \theta_l)\right]  
    \times \exp\left[-\frac{1}{2} \begin{pmatrix} \theta_l - \mu_l \\ \theta_n - \mu_n \end{pmatrix}^T \Sigma^{-1} \begin{pmatrix} \theta_l - \mu_l \\ \theta_n - \mu_n \end{pmatrix} \right]\,.
\end{align*}
\end{widetext}
To marginalize over $\theta_l$, we further define
\be 
\begin{split}
    A &= (\Sigma_{ll} - \Sigma_{ln} \Sigma_{nn}^{-1} \Sigma_{ln}^T)^{-1} = \Sigma_{l|n}^{-1}, \quad X = X(\theta_n) \\
    B &= -A \Sigma_{ln} \Sigma_{nn}^{-1} \\
    D &= \Sigma_{nn}^{-1} + \Sigma_{nn}^{-1} \Sigma_{ln}^T A \Sigma_{ln} \Sigma_{nn}^{-1}
\end{split}
\ee
With some rearrangements, the exponent of the likelihood function becomes a quadratic form in $\theta_l$:
\be 
\begin{split}
    &-\frac{1}{2} \big[ \theta_l^T (X^T C^{-1} X + A) \theta_l - 2 \theta_l^T (X^T C^{-1} (d - m(\theta_n)) \\
    &+ A \mu_{l | n}) \big] + \text{terms without } \theta_l\,.
    \end{split}
\ee
Integrating over $\theta_l$ hence yields the marginalized likelihood function:
\be\begin{split}
    & \mathcal{L}(\theta_n) \propto \frac{1}{\sqrt{\det(X^T C^{-1} X + A)\det(\Sigma_{l|n})\det(\Sigma_{nn})}} \\
    &\times
    \exp\Big[ -\frac{1}{2}(d - m(\theta_n) - X \mu_{l | n})^T \times \\
   & C_{\rm marg}^{-1} (d - m(\theta_n) - X \mu_{l | n}) 
    - \frac{1}{2} (\theta_n - \mu_n)^T D (\theta_n - \mu_n) \Big]\,,
    \end{split}
\ee
where
\be 
    C_{\rm marg}^{-1} = C^{-1} - C^{-1} X (X^T C^{-1} X + A)^{-1} X^T C^{-1}\,.
\ee

\subsection{Analytical Marginalization with a Gaussian Mixture Prior}
We now generalize the analytical marginalization to the case where the prior over parameters is not a single multivariate Gaussian, but a Gaussian Mixture Model (GMM). This means that the model remains
\be 
    d = m(\theta_n) + X(\theta_n) \theta_l
\ee 
where $\theta_n$ are nonlinear parameters and $\theta_l$ are linear nuisance parameters, but now the prior over all parameters is a weighted mixture of multivariate Gaussians:
\be 
    p(\theta) = \sum_{k=1}^K w_k \, \mathcal{N}(\theta; \mu^{(k)}, \Sigma^{(k)})
\ee 
with component means and covariances:
\begin{align}
    \mu^{(k)} = \begin{pmatrix} \mu_l^{(k)} \\ \mu_n^{(k)} \end{pmatrix}, \quad
    \Sigma^{(k)} = \begin{pmatrix} \Sigma_{ll}^{(k)} & \Sigma_{ln}^{(k)} \\
    (\Sigma_{ln}^{(k)})^T & \Sigma_{nn}^{(k)} \end{pmatrix}
\end{align}
Since this is just a sum of Gaussians, we can still analytically marginalize over the linear parameters $\theta_l$ and this can be done for each Gaussian component individually. For component $k$, the conditional distribution of $\theta_l$ given $\theta_n$ is:
\begin{align}
    \mu_{l|n}^{(k)} &= \mu_l^{(k)} + \Sigma_{ln}^{(k)} (\Sigma_{nn}^{(k)})^{-1} (\theta_n - \mu_n^{(k)}) \\
    \Sigma_{l|n}^{(k)} &= \Sigma_{ll}^{(k)} - \Sigma_{ln}^{(k)} (\Sigma_{nn}^{(k)})^{-1} (\Sigma_{ln}^{(k)})^T
\end{align}
Similarly, we can define
\begin{align}
    A^{(k)} &= (\Sigma_{l|n}^{(k)})^{-1}, \quad X = X(\theta_n)\,, \\
    B^{(k)} &= -A^{(k)} \Sigma_{ln}^{(k)} (\Sigma_{nn}^{(k)})^{-1}\,, \\
    D^{(k)} &= (\Sigma_{nn}^{(k)})^{-1} + (\Sigma_{nn}^{(k)})^{-1} (\Sigma_{ln}^{(k)})^T A^{(k)} \Sigma_{ln}^{(k)} (\Sigma_{nn}^{(k)})^{-1}\,,
\end{align}
and write the marginalized likelihood contribution from component $k$ as
\begin{widetext}
\be \begin{split}
    \mathcal{L}_k(\theta_n) \propto &\frac{1}{\sqrt{\det(X^T C^{-1} X + A^{(k)})\det(\Sigma_{l|n}^{(k)})\det(\Sigma_{nn}^{(k)})}} \\
    & \times \exp\left[ -\frac{1}{2}(d - m(\theta_n) - X \mu_{l|n}^{(k)})^T C_{\text{marg}}^{-1} (d - m(\theta_n) - X \mu_{l|n}^{(k)}) \right. 
   \left. - \frac{1}{2} (\theta_n - \mu_n^{(k)})^T D^{(k)} (\theta_n - \mu_n^{(k)}) \right]  \,,
   \end{split}
\ee
\end{widetext}
with
\be 
    C_{\text{marg}}^{-1} = C^{-1} - C^{-1} X (X^T C^{-1} X + A^{(k)})^{-1} X^T C^{-1}\,.
\ee
Therefore, the full marginalized likelihood becomes a mixture of the component-wise marginalized likelihoods, weighted by these conditional weights:
\be 
    \mathcal{L}_{\text{marg}}(\theta_n) = \sum_{k=1}^K w_k \, \mathcal{L}_k(\theta_n)
\ee
Note that the marginalized likelihood is obtained by integrating out $\theta_l$ from the full joint distribution, which already includes the prior. As a result, the original mixture weights $w_k$ remain unchanged. By contrast, if we consider only the conditional distribution $p(\theta_l \mid \theta_n)$, the weights must be renormalized:
\be 
\begin{split}
&p(\theta_l \mid \theta_n)=
\\
=& \sum_{k=1}^{K}
   \frac{w_k\,\mathcal{N}\!\bigl(\theta_n \,\big|\, \mu_n^{(k)},\Sigma_{nn}^{(k)}\bigr)}
        {\sum_{j=1}^{K} w_j\,\mathcal{N}\!\bigl(\theta_n \,\big|\, \mu_n^{(j)},\Sigma_{nn}^{(j)}\bigr)}
   \,\mathcal{N}\!\bigl(\theta_l \,\big|\, \mu_{l|n}^{(k)},\Sigma_{l|n}^{(k)}\bigr) \\
= & \sum_{k=1}^{K}
   w_k'\,\mathcal{N}\!\bigl(\theta_l \,\big|\, \mu_{l|n}^{(k)},\Sigma_{l|n}^{(k)}\bigr),
   \end{split}
\ee
where $\mathcal{N}(\cdot \mid \mu, \Sigma)$ denotes a Gaussian distribution with mean $\mu$ and covariance $\Sigma$. In our case, we evaluate the integral $\int\!d\theta_l \, p(\theta_n,\theta_l)
= \int\!d\theta_l \, p(\theta_l\mid\theta_n)\,p(\theta_n)$. The additional factor of $p(\theta_n)$ cancels the denominator in $w_k'$, and the remaining numerator is simply the prior for $\theta_n$, which is already accounted for in the marginalized likelihood.

\begin{figure*}
    \includegraphics[width=1.0\linewidth]{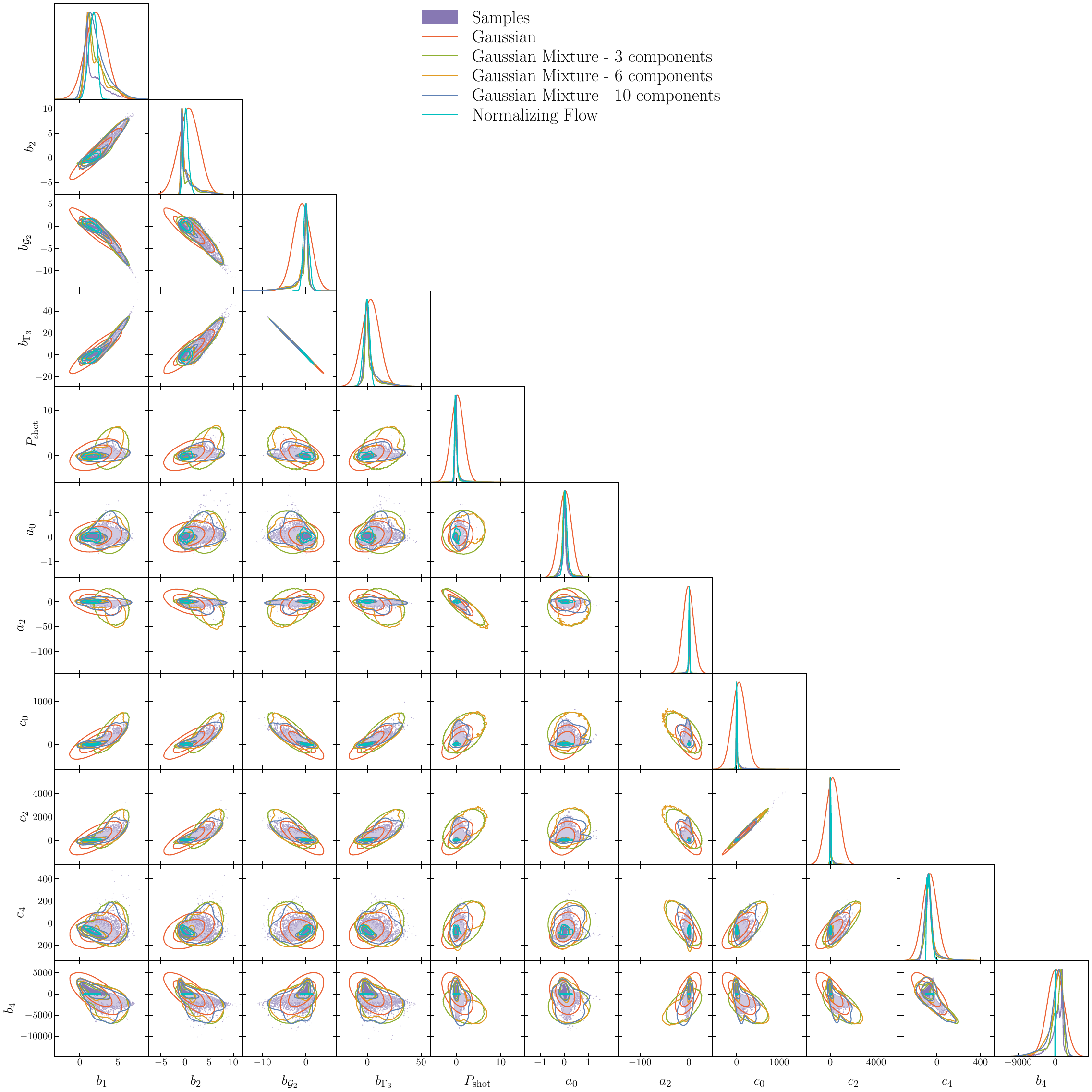}
    \caption{Triangle plot comparing the simulation-based prior (purple filled contours) with several analytic density approximations: a single multivariate Gaussian (orange),  Gaussian mixture models (GMMs) with 3 (green), 6 (yellow), and 10 (blue) components, and prior trained with a normalizing flow (light blue). The plot shows the 68\% and 95\% credible regions for all pairs of parameters used in the posterior sampling. Increasing the number of GMM components progressively improves the agreement with the sample-based data points, capturing non-Gaussian features such as skewness and heavy tails in the distribution.}
    \label{fig:prior_triangle_plot}
\end{figure*}

Fig.~\ref{fig:prior_triangle_plot} compares prior samples 
from Ref.~\cite{Ivanov:2024xgb}
to several analytic approximations of the prior distribution, including a single multivariate Gaussian (orange), Gaussian Mixture Models (GMMs) with 3 (green), 6 (yellow), and 10 (blue) components, and a normalizing flow-based model
used in Ref.~\cite{Ivanov:2024xgb} (light blue). The figure shows that the single Gaussian significantly overestimates the marginal variances for many parameters, failing to capture the heavy tails and skewness inherent in the simulation-based prior. This underscores the limitation of using a single Gaussian for approximating complex, non-Gaussian priors—particularly when analytic marginalization over linearly entering parameters is desired to accelerate posterior sampling.

Among the tested approximations, the GMMs provide a substantially better fit than both the single Gaussian and the normalizing flow. They more accurately capture the skewness, kurtosis, and multimodal features of the prior distribution, even with a modest number of components. In contrast, while normalizing flows offer a flexible, high-capacity modeling approach, their performance suffers when training data is limited. This results in poor representation of the tails and misalignment in the locations of distributional peaks.

The GMMs are trained using the Expectation-Maximization (EM) algorithm to estimate the means and covariances of each component. All parameters are standardized prior to fitting to ensure consistent scaling across dimensions, and a regularization diagonal 
term of $10^{-4}$ is added to the covariance matrices to maintain numerical stability during inversion. To reduce sensitivity to local optima, we perform 100 random initializations and retain the model with the highest log-likelihood.

\section{Datasets}\label{sec:datasets}
In our main analysis, we consider joint analysis of four different datasets: BOSS DR12 galaxy clustering, Planck 2018, Pantheon$+$, and DESI DR2 BAO. We will present the detail of each experiment here and explain the parameters varied in this section.

\paragraph{BOSS DR12 Galaxy Clustering.}
The BOSS DR12 galaxy survey~\cite{BOSS:2012dmf,BOSS:2016wmc} contains galaxies observed in two disjoint regions over the sky, denoted as the northern and southern galactic caps (NGC and SGC). In each two regions, the we have mixed samples from two different sets: CMASS and LOWZ samples, and we divide the redshift range into two slices: $0.2<z<0.5$ and $0.5<z<0.75$, denoted as $z_1$ and $z_3$. In total, we will have four different patches of the sky: $[\text{NGC},\,\text{SGC}]\times[z_1,\,z_3]$, and the combination NGC $\times$ $z_3$ has the largest survey volume. 

To perform the full-shape (FS) analysis, we use four different statistics: redshift-space power spectrum multipoles $P_{\ell}(k)$ with $\ell=0,2,4$, real-space power spectrum $Q_0$~\cite{Ivanov:2021fbu}, 
bispectrum monopole $B_0(k_1,k_2,k_3)$~\cite{Philcox:2021kcw} modeled at the EFT tree-level~\cite{Ivanov:2021kcd}, and BAO parameters $[\alpha_\|,\,\alpha_\perp]$
from the post-reconstucted
power spectra~\cite{Philcox:2020vvt}.
For the scale cut of each spectrum, we have $k_{P_\ell}\in[0.01, 0.20]$ $h/$Mpc, $k_{Q_0}\in[0.20, 0.40]$ $h/$Mpc, and $k_{B_0}\in[0.01, 0.08]$ $h/$Mpc validated in~\cite{Ivanov:2021fbu,Philcox:2021kcw,Chudaykin:2024wlw}. We assume a Gaussian likelihood and the covariance is obtained from the 2048 MultiDark Patchy mocks~\cite{BOSS:2016wmc}. These statistics are estimated with the windowless approach detailed in Ref.~\cite{Philcox:2020vbm,Philcox:2021ukg} and are publicly available~\cite{Philcox:2021kcw}\footnote{\url{https://github.com/oliverphilcox/full_shape_likelihoods}}. 

Beside the cosmological parameter, we vary the three bias parameters, $b_1$, $b_2$, and $b_{\mathcal{G}_2}$, explicitly for each chunk of the sky. The rest of the nuisance parameters that enter linearly are marginalized analytically during the MCMC sampling, which includes $\{b_{\Gamma_3}, P_{\rm shot}, a_0, a_2, c_{s,0}, c_{s,2}, c_{s,4}, b_4\}$ for the power spectrum, and additional $\{c_1, B_{\rm shot}\}$ for the bispectrum. Note that for the latter two we use the standard
conservative Gaussian priors.
We will consider two different prior distribution for these parameters. The first one is the simulation-based prior as discussed in Section~\ref{sec:method}, while the second one is the same conservative prior used in Ref.~\cite{Philcox:2021kcw}. This conservative prior is used to set as a baseline for our analysis. We also set the fiducial galaxy number density as $\bar{n}=3\times10^4~[h^{-1}\text{Mpc}]^3$.

Notably, 
the full-shape 
BOSS (BOSS-FS) results~\cite{Philcox:2021kcw} provide constraints whose errors are
comparable to DESI DR1 full-shape (DESI DR1-FS) data, cf.~\cite{DESI:2024jis}.
Once the simulation-based 
priors are used, the BOSS constraints 
are nominally stronger than 
DESI DR1-FS~\cite{Ivanov:2024xgb}, 
which justifies the use of this dataset 
over the DESI DR1 in our analysis. 
We use the implementation of the simulation-based
priors to DESI DR1-FS to future work.

\paragraph{Planck 2018 CMB.}
We use the publicly available Planck 2018 {\tt Plik} likelihood~\cite{Planck:2019nip,Planck:2018nkj,Planck:2018vyg}\footnote{\url{https://pla.esac.esa.int/pla/\#home}} in our analysis. This includes the high-multipole temperature and polarization data (TT, TE, EE) over the range $\ell \approx 30$–$2500$, derived from cross-spectra between multiple frequency channels. For the low multipole range ($\ell < 30$), we include both temperature (TT) and E-mode polarization (EE) data. In addition, we incorporate the reconstructed CMB lensing potential, which provides complementary constraints on the late-time matter distribution.

\paragraph{Pantheon$+$ Type Ia Supernovae.}
Pantheon$+$ Type Ia supernovae (SNe Ia) likelihood provides distance modulus measurements for 1701 light curves from 1550 unique SNe Ia spanning redshifts from $z=0.001$ up to $2.26$~\cite{Scolnic:2021amr,Brout:2022vxf}. As the sample is uncalibrated in absolute terms, we marginalize over the SN Ia absolute magnitude $M$, treating it as a nuisance parameter.

\paragraph{DESI DR2 BAO.}
We use parts of BAO data from DESI DR2~\cite{DESI:2025zgx} that do not overlap with samples obtained from the BOSS galaxy survey, following~\cite{Chen:2024vuf}. In particular, as a conservative choice, we consider measurements with $z>0.75$, corresponding to part of luminous red galaxies (LRG) samples, and all of emission line galaxies (ELG) and quasars (QSO) samples within the survey. We use the publicly available data within the {\tt Cobaya} sampler\footnote{\url{https://github.com/CobayaSampler/cobaya}}.

\section{Results}\label{sec:res}

\begin{figure*}
    \centering
    \includegraphics[width=\linewidth]{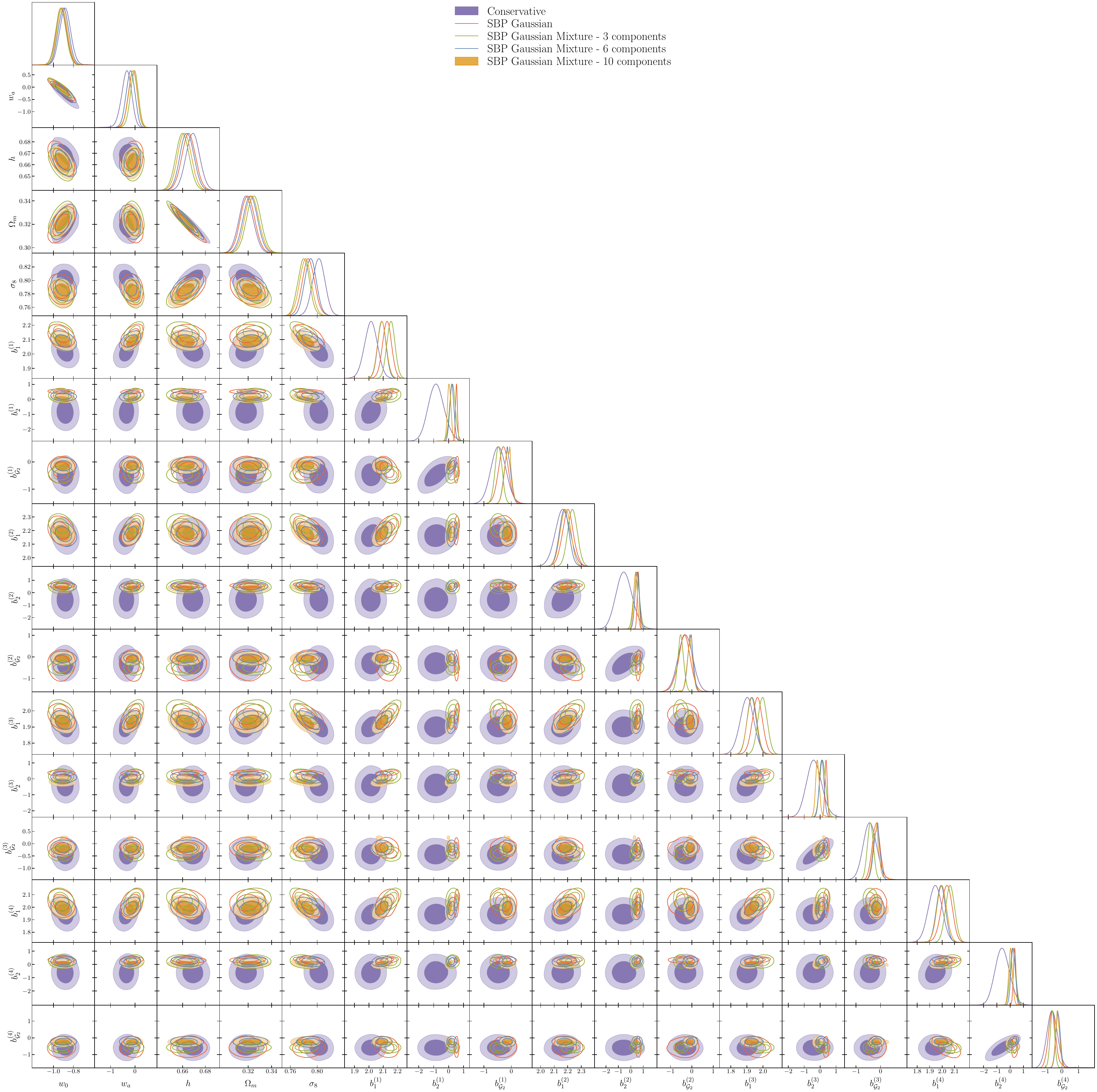}
    \caption{Triangle plot comparing the posterior distribution for selected cosmological and bias parameters under different prior assumptions. The filled purple contours correspond to the posterior obtained using the conservative EFT prior. Overlaid are posterior distributions using simulation-based priors (SBPs) modeled with analytic density approximations: a single multivariate Gaussian (blue), and Gaussian mixture models (GMMs) with 3 (green), 6 (red), and 10 (filled orange) components. The plot shows the 68\% and 95\% credible regions for all parameter pairs. Increasing the number of GMM components progressively improves the ability of the analytic approximation to capture non-Gaussian features in the simulation-based prior, such as skewness and extended tails.}
    \label{fig:posterior_triangle_plot}
\end{figure*}

We present our main result in Table~\ref{tab:bias_cosmo_parameters}
and 
Fig.~\ref{fig:posterior_triangle_plot}, which summarize the inferred values of galaxy bias parameters ($b_1^{(i)}$, $b_2^{(i)}$, and $b_{\mathcal{G}_2}^{(i)}$) across four patches of sky from BOSS galaxy survey, along with constraints on cosmological parameters ($w_0$, $w_a$, $\Omega_m$, $\sigma_8$, and $h$). These constraints are derived from a joint analysis combining BOSS DR12 full shape, Planck 2018, Pantheon+, and DESI DR2 BAO, evaluated under different choices for modeling the prior distribution of nuisance parameters associated with the galaxy survey. In addition to the nuisance parameters specific to each dataset, we vary the following set of cosmological parameters directly: $\{h,\, \ln(10^{10}A_s),\,\omega_{\rm cdm},\,\omega_{\rm b},\,n_s,\,\tau\}$. Throughout, we assume an effective number of neutrino species $N_{\rm eff} = 3.04$ and a single massive neutrino  with mass of 0.06~eV.

\begin{table*}[htbp!]
\scriptsize
\renewcommand{\arraystretch}{1.3}
\begin{tabular}{llcccc}
\multicolumn{6}{c}{\textbf{Experiments: Planck 2018 + BOSS DR12 + Pantheon$+$ + DESI BAO DR2}} \\
\hline
Methods to Approximate Prior & Parameter & Bin 1 & Bin 2 & Bin 3 & Bin 4 \\
\hline
\multirow{8}{*}{Conservative Prior}
& $b_1^{(i)}$ & $2.02^{+0.046}_{-0.0455}$ & $2.16^{+0.0573}_{-0.0548}$ & $1.90^{+0.0436}_{-0.0431}$ & $1.95^{+0.0563}_{-0.0550}$ \\
& $b_2^{(i)}$ & $-0.819^{+0.513}_{-0.572}$ & $-0.524^{+0.630}_{-0.695}$ & $-0.372^{+0.466}_{-0.499}$ & $-0.580^{+0.535}_{-0.579}$ \\
& $b_{\mathcal{G}_2}^{(i)}$ & $-0.471^{+0.279}_{-0.288}$ & $-0.300^{+0.343}_{-0.349}$ & $-0.436^{+0.274}_{-0.278}$ & $-0.583^{+0.323}_{-0.325}$ \\
& $w_0$ & \multicolumn{4}{c}{$-0.884^{+0.0557}_{-0.0560}$} \\
& $w_a$ & \multicolumn{4}{c}{$-0.342^{+0.220}_{-0.191}$} \\
& $\Omega_m$ & \multicolumn{4}{c}{$0.318^{+0.00595}_{-0.00635}$} \\
& $\sigma_8$ & \multicolumn{4}{c}{$0.802^{+0.00904}_{-0.00905}$} \\
& $h$ & \multicolumn{4}{c}{$0.669^{+0.00613}_{-0.00612}$} \\
\hline

\multirow{8}{*}{Gaussian} 
& $b_1^{(i)}$ & $2.13^{+0.0309}_{-0.0318}$ & $2.20^{+0.0397}_{-0.0397}$ & $1.97^{+0.0309}_{-0.0319}$ & $2.04^{+0.0427}_{-0.0416}$ \\
& $b_2^{(i)}$ & $0.523^{+0.0622}_{-0.0621}$ & $0.602^{+0.0833}_{-0.0849}$ & $0.370^{+0.0751}_{-0.0750}$ & $0.342^{+0.0920}_{-0.0926}$ \\
& $b_{\mathcal{G}_2}^{(i)}$ & $-0.301^{+0.190}_{-0.179}$ & $-0.345^{+0.268}_{-0.250}$ & $-0.189^{+0.166}_{-0.156}$ & $-0.602^{+0.223}_{-0.203}$ \\
& $w_0$ & \multicolumn{4}{c}{$-0.922^{+0.0533}_{-0.0544}$} \\
& $w_a$ & \multicolumn{4}{c}{$-0.0957^{+0.189}_{-0.171}$} \\
& $\Omega_m$ & \multicolumn{4}{c}{$0.320^{+0.00607}_{-0.00636}$} \\
& $\sigma_8$ & \multicolumn{4}{c}{$0.787^{+0.00902}_{-0.00920}$} \\
& $h$ & \multicolumn{4}{c}{$0.665^{+0.00615}_{-0.00618}$} \\
\hline

\multirow{8}{*}{Gaussian Mixture, GMM3} 
& $b_1^{(i)}$ & $2.15^{+0.0284}_{-0.0288}$ & $2.24^{+0.0354}_{-0.0370}$ & $2.00^{+0.0285}_{-0.0294}$ & $2.06^{+0.0376}_{-0.0355}$ \\
& $b_2^{(i)}$ & $0.257^{+0.185}_{-0.198}$ & $0.461^{+0.227}_{-0.248}$ & $0.158^{+0.185}_{-0.202}$ & $0.215^{+0.207}_{-0.218}$ \\
& $b_{\mathcal{G}_2}^{(i)}$ & $-0.468^{+0.131}_{-0.133}$ & $-0.514^{+0.140}_{-0.141}$ & $-0.414^{+0.122}_{-0.127}$ & $-0.574^{+0.133}_{-0.134}$ \\
& $w_0$ & \multicolumn{4}{c}{$-0.923^{+0.0535}_{-0.0504}$} \\
& $w_a$ & \multicolumn{4}{c}{$-0.0302^{+0.174}_{-0.161}$} \\
& $\Omega_m$ & \multicolumn{4}{c}{$0.325^{+0.00615}_{-0.00615}$} \\
& $\sigma_8$ & \multicolumn{4}{c}{$0.780^{+0.00863}_{-0.00860}$} \\
& $h$ & \multicolumn{4}{c}{$0.660^{+0.00577}_{-0.00610}$} \\
\hline

\multirow{8}{*}{Gaussian Mixture, GMM6} 
& $b_1^{(i)}$ & $2.09^{+0.0290}_{-0.0296}$ & $2.17^{+0.0373}_{-0.0380}$ & $1.93^{+0.0291}_{-0.0300}$ & $1.99^{+0.0376}_{-0.0377}$ \\
& $b_2^{(i)}$ & $0.213^{+0.142}_{-0.145}$ & $0.513^{+0.176}_{-0.177}$ & $0.0952^{+0.149}_{-0.149}$ & $0.234^{+0.168}_{-0.176}$ \\
& $b_{\mathcal{G}_2}^{(i)}$ & $-0.143^{+0.127}_{-0.127}$ & $-0.0694^{+0.144}_{-0.135}$ & $-0.158^{+0.130}_{-0.129}$ & $-0.231^{+0.137}_{-0.137}$ \\
& $w_0$ & \multicolumn{4}{c}{$-0.900^{+0.0523}_{-0.0548}$} \\
& $w_a$ & \multicolumn{4}{c}{$-0.184^{+0.192}_{-0.177}$} \\
& $\Omega_m$ & \multicolumn{4}{c}{$0.322^{+0.00606}_{-0.00628}$} \\
& $\sigma_8$ & \multicolumn{4}{c}{$0.790^{+0.00873}_{-0.00889}$} \\
& $h$ & \multicolumn{4}{c}{$0.664^{+0.00597}_{-0.00617}$} \\
\hline

\multirow{8}{*}{Gaussian Mixture, GMM10} 
& $b_1^{(i)}$ & $2.09^{+0.0301}_{-0.0317}$ & $2.18^{+0.0399}_{-0.0399}$ & $1.93^{+0.0305}_{-0.0311}$ & $2.00^{+0.0395}_{-0.0405}$ \\
& $b_2^{(i)}$ & $0.0421^{+0.116}_{-0.124}$ & $0.378^{+0.151}_{-0.156}$ & $-0.154^{+0.115}_{-0.146}$ & $0.0332^{+0.145}_{-0.153}$ \\
& $b_{\mathcal{G}_2}^{(i)}$ & $-0.129^{+0.128}_{-0.125}$ & $-0.0921^{+0.145}_{-0.133}$ & $-0.183^{+0.120}_{-0.149}$ & $-0.255^{+0.134}_{-0.141}$ \\
& $w_0$ & \multicolumn{4}{c}{$-0.911^{+0.0519}_{-0.0549}$} \\
& $w_a$ & \multicolumn{4}{c}{$-0.0942^{+0.188}_{-0.170}$} \\
& $\Omega_m$ & \multicolumn{4}{c}{$0.323^{+0.00607}_{-0.00633}$} \\
& $\sigma_8$ & \multicolumn{4}{c}{$0.782^{+0.00889}_{-0.00894}$} \\
& $h$ & \multicolumn{4}{c}{$0.662^{+0.00601}_{-0.00611}$} \\
\hline

\end{tabular}
\caption{Constraints on linear, quadratic, and tidal bias parameters ($b_1^{(i)}$, $b_2^{(i)}$, $b_{G2}^{(i)}$) across four redshift bins, along with cosmological parameters ($w_0$, $w_a$, $\Omega_m$, $\sigma_8$, $h$), obtained from joint analyses using Planck 2018, BOSS, Pantheon+, and DESI BAO DR2. Results are shown for different prior modeling choices, including the conservative prior and the SBI prior approximated with Gaussian and Gaussian mixture approaches. We ensure that the Gelman-Rubin statistics satisfies $R<0.01$ for all parameters.}
\label{tab:bias_cosmo_parameters}
\end{table*}

\begin{table}[htbp!]
\begin{tabular}{lcl}
\hline
Method & FoM ($w_0$--$w_a$) \\
\hline
Conservative & 206.06 \\
Gaussian & 242.70 \\
GMM (3 components) & 268.21 \\
GMM (6 components) & 240.54 \\
GMM (10 components) & 248.81 \\
\hline
\end{tabular}
\caption{Figure of Merit (FoM) for $w_0$ and $w_a$ derived from joint analyses using Planck 2018, BOSS, Pantheon+, and DESI BAO DR2 with different choices of prior distribution on EFT parameters. The FoM is defined as $\mathrm{FoM} = 1 / \sqrt{\det \mathrm{Cov}(w_0, w_a)}$, where $\mathrm{Cov}(w_0, w_a)$ is the marginalized $2\times2$ covariance matrix. “Gaussian” indicates a Gaussian approximation to the prior; “GMM” denotes that the prior is approximated by the Gaussian Mixture Model. Except the conservative one, all cases incorporate simulation-based priors.\label{tab:fom}}
\end{table}

\begin{figure*}[htbp!]
    \includegraphics[width=\linewidth]{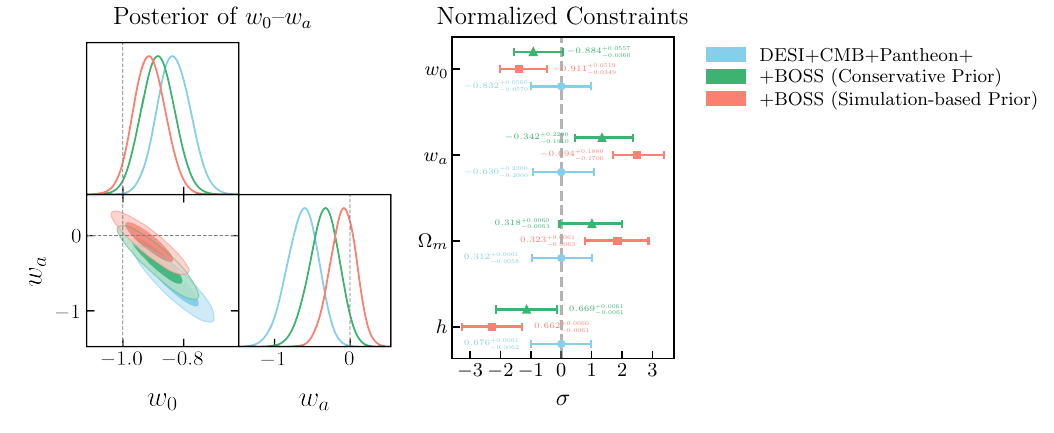}
    \caption{Constraints on cosmological parameters under different analysis configurations. The left panel shows the 2D 68\% and 95\% confidence regions in the $w_0$–$w_a$ plane, while the right panel presents the normalized 1D confidence intervals for $w_0$, $w_a$, $\Omega_m$, and $h$. The baseline result (DESI+CMB+Pantheon$+$; blue) combines DESI DR2 BAO measurements, Planck 2018 CMB data, and Pantheon$+$ supernovae, and is used to normalize both the central values and uncertainty widths of the other two configurations. Note that the CMB dataset used here differs from that in Ref.~\cite{DESI:2025zgx}, which includes additional CMB experiments, but the result is consistent with using Planck 2018 data alone. The other two results incorporate the BOSS dataset using either a simulation-based prior (orange) or a conservative prior (green). Further details on the datasets and prior choices can be found in Section~\ref{sec:datasets}. Importantly, when BOSS data is added, part of the DESI DR2 BAO sample is removed to avoid double-counting information. }
\label{fig:1d_cosmo_posterior_comparison}
\end{figure*}

Finally, let us compare our cosmological constraints with the results from Ref.~\cite{DESI:2025zgx} that is the same as us without the inclusion of BOSS galaxy clustering data. Fig.~\ref{fig:1d_cosmo_posterior_comparison} shows the comparison on the cosmological parameters, focusing on $w_0$, $w_a$, $\Omega_m$ and $h$. We found that our result based on GMM10 prefers $w_0\rightarrow-1$ and $w_a\rightarrow0$ for the dark energy model.  

Table~\ref{tab:bias_cosmo_parameters}
also compares the impact of various prior approximations: a conservative Gaussian prior, a single multivariate Gaussian fit to simulation-based samples, and GMM with 3, 6, and 10 components (hereafter GMM3, GMM6, and GMM10, respectively). The conservative Gaussian prior, widely adopted in earlier EFT-based studies~\cite{Philcox:2021kcw,Ivanov:2023qzb,Chen:2024vuf}, offers analytic tractability but lacks non-perturbative information from small-scale simulations, potentially yielding overly cautious estimates for weakly constrained nuisance parameters.

As discussed in Section~\ref{sec:method}, in order to speed up the sampling process with priors motivated from HOD simulations, we adopt GMM so that we can perform analytic marginalization. We start from a single multivariate Gaussian approximation and then increase the number of components to improve flexibility. Among all the bias parameters, we find that using a single Gaussian to approximate the simulation-based prior causes a notable shift--greater than $2\sigma$--from the conservative prior prediction for both $b_1^{(i)}$ and $b_2^{(i)}$. In contrast, the posteriors of $b_{\mathcal{G}_2}^{(i)}$ remain broadly consistent with those from the conservative prior. The significant shifts of bias parameters are the consequence of the single Gaussian's poor ability to capture the non-Gaussian features (such as skewness and heavy tails) present in the simulation-based prior. As shown in Fig.~\ref{fig:prior_triangle_plot}, the peaks of the single-Gaussian fit for parameters like $b_1$ and $b_2$ are clearly misaligned with the sample distribution.

As we increase the number of components in the GMM, the inferred bias parameters become increasingly more consistent with the original simulation-based prior. The GMM3 and GMM6 models already show noticeable improvement in both the central values and the reduction of variances, while GMM10 yields the most flexible and faithful approximation. Importantly, the results from GMM6 and GMM10 agree with each other within the two $1\sigma$ confidence interval, indicating that the mixture model has effectively converged, especially for $b_1$. 

Although the single Gaussian performs poorly for some bias parameters, its predictions for cosmological parameters remain broadly consistent with those from the GMM10 model. For example, both of them prefer a lower $\sigma_8$ compared with the one obtained from the conservative prior at the level of $1\sigma$, similar to the result reported in Ref.~\cite{Ivanov:2024xgb}. This indicates that, in this case, the cosmological inference is relatively robust to the specific shape of the nuisance prior--likely due to limited degeneracies between cosmological and bias parameters as we have multiple observations combined here. However, relying solely on single Gaussian might be inaccurate given
the errorbars of upcoming data releases. 
We further test different approximations
in Appendix~\ref{eq:app}, where 
we show that the Gaussian approximation 
leads to wider contours than the normalizing 
flows or the GMM. We note, however, 
that the Gaussian approximation may be 
considered a conservative choice
given that it produces 
larger posteriors than the other modeling
options.

All MCMC chains pass the Gelman-Rubin convergence criterion ($R-1<0.01$) for all parameters. These results demonstrate that simulation-based priors, when approximated using GMMs, can be effectively and efficiently integrated into usual full shape analysis of galaxy surveys. GMMs offer a powerful tool other than normalizaing flow for handling high-dimensional, non-Gaussian nuisance prior structures without sacrificing analytic marginalizability or the robustness of cosmological parameter estimations. 

To further quantify the impact of prior modeling on dark energy constraints, we present in Table~\ref{tab:fom} the Figure of Merit (FoM) for $w_0-w_a$ plane, defined as ${\rm FoM}=1/\sqrt{{\rm det}\,{\rm Cov}(w_0, w_a)}$~\cite{Albrecht:2006um}. This metric captures the inverse area of the joint confidence ellipse in the $w_0$-$w_a$ plane and reflects the precision of dark energy constraints. Among all cases, the GMM10 provides the most faithful approximation of the simulation-based prior and should be considered the most accurate baseline. It balances flexibility, numerical stability, and analytic marginalizability, and its results are fully converged with respect to prior modeling.

Compared to GMM10, the conservative prior leads to a 20\% reduction in the FoM, indicating that using overly broad and uninformative priors discards meaningful information about nuisance parameter correlations captured by simulations. The single Gaussian improves the FoM relative to the conservative case but still slightly underperforms compared to GMM10. Interestingly, GMM3 achieves a slightly higher FoM than GMM10, likely due to small numerical artifacts or implicit regularization from using fewer components; however, this comes at the cost of lower fidelity to the true prior structure, so it does not indicate a more accurate constraint.

Taken together, these results show that while less flexible models (e.g., GMM3 or a single Gaussian) may yield superficially tighter constraints, they can also introduce biases or misrepresentations of parameter uncertainties. The GMM10-based result offers the most robust and interpretable constraint on dark energy, and should be regarded as the reference for drawing physical conclusions. Our final results 
in terms of the constraints 
on the $w_0,w_a,\Omega_m$
and $h$ parameters
are displayed in Fig.~\ref{fig:1d_cosmo_posterior_comparison}.

Fig.~\ref{fig:posterior_triangle_plot} shows the 68\% and 95\% credible regions for the posterior distributions under different prior assumptions. In particular, we highlight the filled purple and orange contours, corresponding to the conservative prior and the simulation-based prior approximated with GMM10, respectively. The figure demonstrates that while both priors yield consistent central values for cosmological parameters, the GMM10-based prior leads to visibly tighter constraints, underscoring the advantage of incorporating simulation-informed nuisance priors in dark energy analyses.

\section{Discussion}
\label{sec:disc}

Let us start our discussion 
with the physical interpretation
of our results. 
First, we show that the 
CMB+DESI BAO+ Pantheon+ SNe
evidence for dynamical 
dark energy is significantly 
reduced once this dataset is combined
with the full-shape galaxy power 
spectrum and bispectrum data 
from the BOSS survey. 
A similar conclusion was drawn before
in Ref.~\cite{Chen:2024vuf},
although our result is stronger
because Ref.~\cite{Chen:2024vuf}
used DESI BAO DR1, while here we 
use DESI BAO DR2, which is more constraining,
and hence implying a stronger evidence 
for dynamical dark energy in the absence
of BOSS-EFT-FS. As we can see from Fig.~\ref{fig:1d_cosmo_posterior_comparison},
the reduction of preference for 
the $w_0w_a$ model in our dataset
is accompanied by an upper shift 
of $\Omega_m$, which is consistent 
with the observation that 
the evidence for the 
$w_0w_a$ model arises due to the 
tension between DESI and Planck CMB
at the level of this parameter~\cite{DESI:2025fii,Tang:2024lmo,Sailer:2025lxj}. 

Second, we show that the addition of the 
simulation-based priors (SBPs) calibrated 
at the field level to the BOSS full-shape analysis leads to a significant
20\% reduction of the posterior area 
in the $w_0-w_a$ plane. In addition, 
it shifts the $w_0-w_a$ posterior
contour higher up, which 
shifts the 1d marginalized posteriors
for $w_0$ and $w_a$ closer to their
$\Lambda$CDM values $0$ and $-1$, respectively. 
The $w_0$ and $w_a$ 
marginals of the SBP analysis 
are consistent with the cosmological constant
within $95\%$ CL.

All in all, our analysis 
thus shows that the preference 
for dynamical dark energy 
in the DESI BAO+CMB+SNe dataset
significantly reduces upon addition
of the BOSS EFT-based full-shape
power spectrum and bispectrum data, 
and even more so when the latter 
is enhanced with the simulation-based
priors.

We note that our results are
different from Ref.~\cite{DESI:2025wzd},
which analyzed DESI DR2 BAO+Pantheon+ SNe+Planck CMB + DESI DR1-FS and did not find a noticeable reduction
of the $w_0-w_a$ posterior area as a result
of adding their analog of simulation-based
priors. While their analysis was based
on DESI DR1-FS as opposed to 
BOSS FS which we use here, the constraining
power of these data sets is approximately similar,
so we do not expect the difference to be 
driven by the choice of the dataset. 
We believe that we find significantly
stronger constraints because our 
priors are calibrated at the field level, 
while the priors of~\cite{DESI:2025wzd}
are calibrated at 
the power spectrum level. As discussed
in~\cite{Ivanov:2024xgb}, this mode
of EFT parameter measurements does not
allow for a degeneracy breaking,
and hence results in wider,
more noisy prior distribution,
which diminishes the eventual 
gains in parameter constraints. 
It will be interesting to re-analyze
the 
DESI DR2 BAO+Pantheon+ SNe+Planck CMB + DESI DR1-FS
with the field-level simulation-based priors. 
We leave this for future work. 

On the technical side, 
our work is the first application
of the mixed Gaussian models
to fit the distribution 
of EFT parameters from 
N-body-HOD simulations. 
We show that such modeling 
is accurate, reliable, and offers 
a significant gain in efficiency
w.r.t. the normalizing flows 
utilized previously~\cite{Ivanov:2024hgq,Ivanov:2024xgb}.
In particular, we have shown that the 
results of our analysis 
quickly converge 
with a number of 
Gaussian distributions 
in the mixture model used to 
fit the EFT priors.
Specifically, we have found 
that the figure of merit changes 
by less than 4\% when increasing the number 
of Gaussian mixture components from 6 to 10.
It will be interesting to explore other 
models for the EFT posterior
distribution
in order to quantify the modeling 
uncertainties better.

Finally, the tools that we have developed 
here 
will make it easier to implement 
simulation-based priors in cosmological analyses. This opens up new possibilities
for beyond-$\Lambda$CDM full-shape analyses along the lines of~\cite{Xu:2021rwg,Rogers:2023ezo,He:2023dbn,He:2023oke,He:2025npy,Ivanov:2020ril,McDonough:2023qcu,Toomey:2024ita},
as well as opportunities for novel EFT-based analyses of the Lyman-$\alpha$ forest~\cite{Ivanov:2023yla,Ivanov:2024jtl,
deBelsunce:2024rvv,Chudaykin:2025gsh,He:2025jwp,Hadzhiyska:2025cvk}, where 
the simulation-based 
priors are necessary for the best performance. We leave all these research 
directions for future investigation.

\paragraph*{Acknowledgments.} 
MI would like to thank 
Anton Chudaykin,
Shi-Fan Chen, and 
Jamie Sullivan
for helpful discussions.
This work was performed in part at Aspen Center for Physics, which is supported by National Science Foundation grant PHY-2210452.

\appendix

\section{Comparison between Gaussian Mixture Approximation and Normalizing Flow}
\label{eq:app}
To assess the quality of our Gaussian Mixture Model (GMM) approximation, we compare its performance against a baseline derived from direct posterior sampling. In this setup, we sample nuisance parameters—those entering linearly into the theoretical power spectrum—using a log-likelihood function informed by a normalizing flow trained on the full posterior. This trained flow serves as a surrogate for the true posterior distribution and allows us to benchmark the GMM approximation under consistent conditions.

\begin{figure*}
    \centering
    \includegraphics[width=\linewidth]{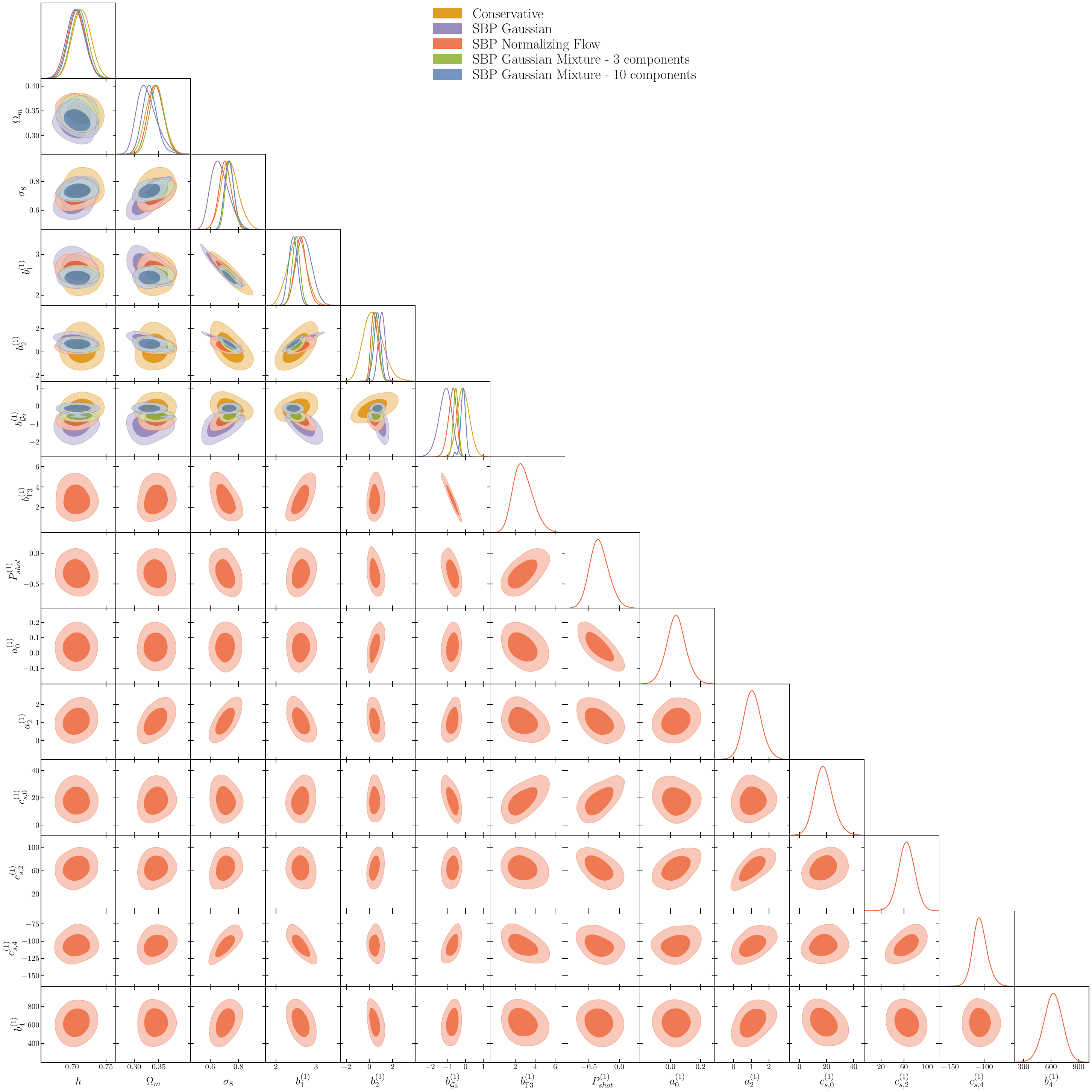}
    \caption{Triangle plot showing marginalized posterior distributions and 2D credible regions for cosmological parameters and bias parameters using a conservative prior (yellow) and three SBP models: Gaussian (purple), Gaussian Mixture with 3 and 10 components (GMM3: green; GMM10: blue), and normalizing flow (red). We can observe better agreement on the constraints of bias parameters between GMM10 and normalizing flow.}
    \label{fig:triangle_gmm_vs_samples_all_posterior_ngc_z3}
\end{figure*}

Fig.~\ref{fig:triangle_gmm_vs_samples_all_posterior_ngc_z3} presents this comparison across three different SBP models and the conservative prior. In this comparison, we use only data of BOSS galaxy clustering from one chunk of sky: NGC $\times$ $z_3$, including $P_\ell$, $Q_0$, $B_0$ and $[\alpha_\|,\alpha_\perp]$. We find that the GMM approximation yields posterior contours that align more closely with those obtained from the normalizing flow baseline than with those from a single Gaussian approximation, demonstrating its effectiveness in capturing key features of the posterior distribution. However, the normalizing flow still tends to produce broader constraints and exhibits noticeable misalignment in the peak locations relative to the GMM. This behavior is consistent with the earlier observations in Fig.~\ref{fig:prior_triangle_plot}, where the GMM more accurately reproduces the sample distribution—particularly in the tails and the location of the peak—compared to the normalizing flow. Notably, in Fig.~\ref{fig:prior_triangle_plot}, the peak generated by the normalizing flow for the parameter $b_1$ is closer to that of the single Gaussian approximation than to the true sample distribution. As a result, the normalizing flow agrees more closely with the single Gaussian approximation for $b_1$, further illustrating its limitations in accurately modeling the shape of the prior. Finally, we also present the result using the conservative prior, from which we can see that contours from GMM10 mostly fall into the 68\% region of the one from the conservative prior.

\bibliography{ref.bib}

\begin{thebibliography}{99}%
\makeatletter
\providecommand \@ifxundefined [1]{%
 \@ifx{#1\undefined}
}%
\providecommand \@ifnum [1]{%
 \ifnum #1\expandafter \@firstoftwo
 \else \expandafter \@secondoftwo
 \fi
}%
\providecommand \@ifx [1]{%
 \ifx #1\expandafter \@firstoftwo
 \else \expandafter \@secondoftwo
 \fi
}%
\providecommand \natexlab [1]{#1}%
\providecommand \enquote  [1]{``#1''}%
\providecommand \bibnamefont  [1]{#1}%
\providecommand \bibfnamefont [1]{#1}%
\providecommand \citenamefont [1]{#1}%
\providecommand \href@noop [0]{\@secondoftwo}%
\providecommand \href [0]{\begingroup \@sanitize@url \@href}%
\providecommand \@href[1]{\@@startlink{#1}\@@href}%
\providecommand \@@href[1]{\endgroup#1\@@endlink}%
\providecommand \@sanitize@url [0]{\catcode `\\12\catcode `\$12\catcode `\&12\catcode `\#12\catcode `\^12\catcode `\_12\catcode `\%12\relax}%
\providecommand \@@startlink[1]{}%
\providecommand \@@endlink[0]{}%
\providecommand \url  [0]{\begingroup\@sanitize@url \@url }%
\providecommand \@url [1]{\endgroup\@href {#1}{\urlprefix }}%
\providecommand \urlprefix  [0]{URL }%
\providecommand \Eprint [0]{\href }%
\providecommand \doibase [0]{http://dx.doi.org/}%
\providecommand \selectlanguage [0]{\@gobble}%
\providecommand \bibinfo  [0]{\@secondoftwo}%
\providecommand \bibfield  [0]{\@secondoftwo}%
\providecommand \translation [1]{[#1]}%
\providecommand \BibitemOpen [0]{}%
\providecommand \bibitemStop [0]{}%
\providecommand \bibitemNoStop [0]{.\EOS\space}%
\providecommand \EOS [0]{\spacefactor3000\relax}%
\providecommand \BibitemShut  [1]{\csname bibitem#1\endcsname}%
\let\auto@bib@innerbib\@empty
\bibitem [{\citenamefont {Adame}\ \emph {et~al.}(2025)\citenamefont {Adame} \emph {et~al.}}]{DESI:2024mwx}%
  \BibitemOpen
  \bibfield  {author} {\bibinfo {author} {\bibfnamefont {A.~G.}\ \bibnamefont {Adame}} \emph {et~al.} (\bibinfo {collaboration} {DESI}),\ }\href {\doibase 10.1088/1475-7516/2025/02/021} {\bibfield  {journal} {\bibinfo  {journal} {JCAP}\ }\textbf {\bibinfo {volume} {02}},\ \bibinfo {pages} {021} (\bibinfo {year} {2025})},\ \Eprint {http://arxiv.org/abs/2404.03002} {arXiv:2404.03002 [astro-ph.CO]} \BibitemShut {NoStop}%
\bibitem [{\citenamefont {Adame}\ \emph {et~al.}(2024{\natexlab{a}})\citenamefont {Adame} \emph {et~al.}}]{DESI:2024uvr}%
  \BibitemOpen
  \bibfield  {author} {\bibinfo {author} {\bibfnamefont {A.~G.}\ \bibnamefont {Adame}} \emph {et~al.} (\bibinfo {collaboration} {DESI}),\ }\href@noop {} {\  (\bibinfo {year} {2024}{\natexlab{a}})},\ \Eprint {http://arxiv.org/abs/2404.03000} {arXiv:2404.03000 [astro-ph.CO]} \BibitemShut {NoStop}%
\bibitem [{\citenamefont {Abdul~Karim}\ \emph {et~al.}(2025)\citenamefont {Abdul~Karim} \emph {et~al.}}]{DESI:2025zgx}%
  \BibitemOpen
  \bibfield  {author} {\bibinfo {author} {\bibfnamefont {M.}~\bibnamefont {Abdul~Karim}} \emph {et~al.} (\bibinfo {collaboration} {DESI}),\ }\href@noop {} {\  (\bibinfo {year} {2025})},\ \Eprint {http://arxiv.org/abs/2503.14738} {arXiv:2503.14738 [astro-ph.CO]} \BibitemShut {NoStop}%
\bibitem [{\citenamefont {Lodha}\ \emph {et~al.}(2025)\citenamefont {Lodha} \emph {et~al.}}]{DESI:2025fii}%
  \BibitemOpen
  \bibfield  {author} {\bibinfo {author} {\bibfnamefont {K.}~\bibnamefont {Lodha}} \emph {et~al.} (\bibinfo {collaboration} {DESI}),\ }\href@noop {} {\  (\bibinfo {year} {2025})},\ \Eprint {http://arxiv.org/abs/2503.14743} {arXiv:2503.14743 [astro-ph.CO]} \BibitemShut {NoStop}%
\bibitem [{\citenamefont {Laureijs}\ \emph {et~al.}(2011)\citenamefont {Laureijs} \emph {et~al.}}]{Laureijs:2011gra}%
  \BibitemOpen
  \bibfield  {author} {\bibinfo {author} {\bibfnamefont {R.}~\bibnamefont {Laureijs}} \emph {et~al.} (\bibinfo {collaboration} {EUCLID}),\ }\href@noop {} {\  (\bibinfo {year} {2011})},\ \Eprint {http://arxiv.org/abs/1110.3193} {arXiv:1110.3193 [astro-ph.CO]} \BibitemShut {NoStop}%
\bibitem [{\citenamefont {Ivezi\'c}\ \emph {et~al.}(2019)\citenamefont {Ivezi\'c} \emph {et~al.}}]{LSST:2008ijt}%
  \BibitemOpen
  \bibfield  {author} {\bibinfo {author} {\bibfnamefont {v.}~\bibnamefont {Ivezi\'c}} \emph {et~al.} (\bibinfo {collaboration} {LSST}),\ }\href {\doibase 10.3847/1538-4357/ab042c} {\bibfield  {journal} {\bibinfo  {journal} {Astrophys. J.}\ }\textbf {\bibinfo {volume} {873}},\ \bibinfo {pages} {111} (\bibinfo {year} {2019})},\ \Eprint {http://arxiv.org/abs/0805.2366} {arXiv:0805.2366 [astro-ph]} \BibitemShut {NoStop}%
\bibitem [{\citenamefont {Akeson}\ \emph {et~al.}(2019)\citenamefont {Akeson} \emph {et~al.}}]{Akeson:2019biv}%
  \BibitemOpen
  \bibfield  {author} {\bibinfo {author} {\bibfnamefont {R.}~\bibnamefont {Akeson}} \emph {et~al.},\ }\href@noop {} {\  (\bibinfo {year} {2019})},\ \Eprint {http://arxiv.org/abs/1902.05569} {arXiv:1902.05569 [astro-ph.IM]} \BibitemShut {NoStop}%
\bibitem [{\citenamefont {McAlpine}\ \emph {et~al.}(2016)\citenamefont {McAlpine} \emph {et~al.}}]{McAlpine:2015tma}%
  \BibitemOpen
  \bibfield  {author} {\bibinfo {author} {\bibfnamefont {S.}~\bibnamefont {McAlpine}} \emph {et~al.},\ }\href {\doibase 10.1016/j.ascom.2016.02.004} {\bibfield  {journal} {\bibinfo  {journal} {Astron. Comput.}\ }\textbf {\bibinfo {volume} {15}},\ \bibinfo {pages} {72} (\bibinfo {year} {2016})},\ \Eprint {http://arxiv.org/abs/1510.01320} {arXiv:1510.01320 [astro-ph.GA]} \BibitemShut {NoStop}%
\bibitem [{\citenamefont {Springel}\ \emph {et~al.}(2018)\citenamefont {Springel} \emph {et~al.}}]{Springel:2017tpz}%
  \BibitemOpen
  \bibfield  {author} {\bibinfo {author} {\bibfnamefont {V.}~\bibnamefont {Springel}} \emph {et~al.},\ }\href {\doibase 10.1093/mnras/stx3304} {\bibfield  {journal} {\bibinfo  {journal} {Mon. Not. Roy. Astron. Soc.}\ }\textbf {\bibinfo {volume} {475}},\ \bibinfo {pages} {676} (\bibinfo {year} {2018})},\ \Eprint {http://arxiv.org/abs/1707.03397} {arXiv:1707.03397 [astro-ph.GA]} \BibitemShut {NoStop}%
\bibitem [{\citenamefont {Hern\'andez-Aguayo}\ \emph {et~al.}(2023)\citenamefont {Hern\'andez-Aguayo} \emph {et~al.}}]{Hernandez-Aguayo:2022xcl}%
  \BibitemOpen
  \bibfield  {author} {\bibinfo {author} {\bibfnamefont {C.}~\bibnamefont {Hern\'andez-Aguayo}} \emph {et~al.},\ }\href {\doibase 10.1093/mnras/stad1657} {\bibfield  {journal} {\bibinfo  {journal} {Mon. Not. Roy. Astron. Soc.}\ }\textbf {\bibinfo {volume} {524}},\ \bibinfo {pages} {2556} (\bibinfo {year} {2023})},\ \Eprint {http://arxiv.org/abs/2210.10059} {arXiv:2210.10059 [astro-ph.CO]} \BibitemShut {NoStop}%
\bibitem [{\citenamefont {Berlind}\ and\ \citenamefont {Weinberg}(2002)}]{Berlind:2001xk}%
  \BibitemOpen
  \bibfield  {author} {\bibinfo {author} {\bibfnamefont {A.~A.}\ \bibnamefont {Berlind}}\ and\ \bibinfo {author} {\bibfnamefont {D.~H.}\ \bibnamefont {Weinberg}},\ }\href {\doibase 10.1086/341469} {\bibfield  {journal} {\bibinfo  {journal} {Astrophys. J.}\ }\textbf {\bibinfo {volume} {575}},\ \bibinfo {pages} {587} (\bibinfo {year} {2002})},\ \Eprint {http://arxiv.org/abs/astro-ph/0109001} {arXiv:astro-ph/0109001} \BibitemShut {NoStop}%
\bibitem [{\citenamefont {Kravtsov}\ \emph {et~al.}(2004)\citenamefont {Kravtsov}, \citenamefont {Berlind}, \citenamefont {Wechsler}, \citenamefont {Klypin}, \citenamefont {Gottloeber}, \citenamefont {Allgood},\ and\ \citenamefont {Primack}}]{Kravtsov:2003sg}%
  \BibitemOpen
  \bibfield  {author} {\bibinfo {author} {\bibfnamefont {A.~V.}\ \bibnamefont {Kravtsov}}, \bibinfo {author} {\bibfnamefont {A.~A.}\ \bibnamefont {Berlind}}, \bibinfo {author} {\bibfnamefont {R.~H.}\ \bibnamefont {Wechsler}}, \bibinfo {author} {\bibfnamefont {A.~A.}\ \bibnamefont {Klypin}}, \bibinfo {author} {\bibfnamefont {S.}~\bibnamefont {Gottloeber}}, \bibinfo {author} {\bibfnamefont {B.}~\bibnamefont {Allgood}}, \ and\ \bibinfo {author} {\bibfnamefont {J.~R.}\ \bibnamefont {Primack}},\ }\href {\doibase 10.1086/420959} {\bibfield  {journal} {\bibinfo  {journal} {Astrophys. J.}\ }\textbf {\bibinfo {volume} {609}},\ \bibinfo {pages} {35} (\bibinfo {year} {2004})},\ \Eprint {http://arxiv.org/abs/astro-ph/0308519} {arXiv:astro-ph/0308519} \BibitemShut {NoStop}%
\bibitem [{\citenamefont {Zheng}\ \emph {et~al.}(2005)\citenamefont {Zheng}, \citenamefont {Berlind}, \citenamefont {Weinberg}, \citenamefont {Benson}, \citenamefont {Baugh}, \citenamefont {Cole}, \citenamefont {Dave}, \citenamefont {Frenk}, \citenamefont {Katz},\ and\ \citenamefont {Lacey}}]{Zheng:2004id}%
  \BibitemOpen
  \bibfield  {author} {\bibinfo {author} {\bibfnamefont {Z.}~\bibnamefont {Zheng}}, \bibinfo {author} {\bibfnamefont {A.~A.}\ \bibnamefont {Berlind}}, \bibinfo {author} {\bibfnamefont {D.~H.}\ \bibnamefont {Weinberg}}, \bibinfo {author} {\bibfnamefont {A.~J.}\ \bibnamefont {Benson}}, \bibinfo {author} {\bibfnamefont {C.~M.}\ \bibnamefont {Baugh}}, \bibinfo {author} {\bibfnamefont {S.}~\bibnamefont {Cole}}, \bibinfo {author} {\bibfnamefont {R.}~\bibnamefont {Dave}}, \bibinfo {author} {\bibfnamefont {C.~S.}\ \bibnamefont {Frenk}}, \bibinfo {author} {\bibfnamefont {N.}~\bibnamefont {Katz}}, \ and\ \bibinfo {author} {\bibfnamefont {C.~G.}\ \bibnamefont {Lacey}},\ }\href {\doibase 10.1086/466510} {\bibfield  {journal} {\bibinfo  {journal} {Astrophys. J.}\ }\textbf {\bibinfo {volume} {633}},\ \bibinfo {pages} {791} (\bibinfo {year} {2005})},\ \Eprint {http://arxiv.org/abs/astro-ph/0408564} {arXiv:astro-ph/0408564} \BibitemShut {NoStop}%
\bibitem [{\citenamefont {Hearin}\ \emph {et~al.}(2016)\citenamefont {Hearin}, \citenamefont {Zentner}, \citenamefont {van~den Bosch}, \citenamefont {Campbell},\ and\ \citenamefont {Tollerud}}]{Hearin:2015jnf}%
  \BibitemOpen
  \bibfield  {author} {\bibinfo {author} {\bibfnamefont {A.~P.}\ \bibnamefont {Hearin}}, \bibinfo {author} {\bibfnamefont {A.~R.}\ \bibnamefont {Zentner}}, \bibinfo {author} {\bibfnamefont {F.~C.}\ \bibnamefont {van~den Bosch}}, \bibinfo {author} {\bibfnamefont {D.}~\bibnamefont {Campbell}}, \ and\ \bibinfo {author} {\bibfnamefont {E.}~\bibnamefont {Tollerud}},\ }\href {\doibase 10.1093/mnras/stw840} {\bibfield  {journal} {\bibinfo  {journal} {Mon. Not. Roy. Astron. Soc.}\ }\textbf {\bibinfo {volume} {460}},\ \bibinfo {pages} {2552} (\bibinfo {year} {2016})},\ \Eprint {http://arxiv.org/abs/1512.03050} {arXiv:1512.03050 [astro-ph.CO]} \BibitemShut {NoStop}%
\bibitem [{\citenamefont {Wechsler}\ and\ \citenamefont {Tinker}(2018)}]{Wechsler:2018pic}%
  \BibitemOpen
  \bibfield  {author} {\bibinfo {author} {\bibfnamefont {R.~H.}\ \bibnamefont {Wechsler}}\ and\ \bibinfo {author} {\bibfnamefont {J.~L.}\ \bibnamefont {Tinker}},\ }\href {\doibase 10.1146/annurev-astro-081817-051756} {\bibfield  {journal} {\bibinfo  {journal} {Ann. Rev. Astron. Astrophys.}\ }\textbf {\bibinfo {volume} {56}},\ \bibinfo {pages} {435} (\bibinfo {year} {2018})},\ \Eprint {http://arxiv.org/abs/1804.03097} {arXiv:1804.03097 [astro-ph.GA]} \BibitemShut {NoStop}%
\bibitem [{\citenamefont {Krause}\ \emph {et~al.}(2024)\citenamefont {Krause} \emph {et~al.}}]{Beyond-2pt:2024mqz}%
  \BibitemOpen
  \bibfield  {author} {\bibinfo {author} {\bibfnamefont {E.}~\bibnamefont {Krause}} \emph {et~al.} (\bibinfo {collaboration} {Beyond-2pt}),\ }\href@noop {} {\  (\bibinfo {year} {2024})},\ \Eprint {http://arxiv.org/abs/2405.02252} {arXiv:2405.02252 [astro-ph.CO]} \BibitemShut {NoStop}%
\bibitem [{\citenamefont {Alam}\ \emph {et~al.}(2017)\citenamefont {Alam} \emph {et~al.}}]{BOSS:2016wmc}%
  \BibitemOpen
  \bibfield  {author} {\bibinfo {author} {\bibfnamefont {S.}~\bibnamefont {Alam}} \emph {et~al.} (\bibinfo {collaboration} {BOSS}),\ }\href {\doibase 10.1093/mnras/stx721} {\bibfield  {journal} {\bibinfo  {journal} {Mon. Not. Roy. Astron. Soc.}\ }\textbf {\bibinfo {volume} {470}},\ \bibinfo {pages} {2617} (\bibinfo {year} {2017})},\ \Eprint {http://arxiv.org/abs/1607.03155} {arXiv:1607.03155 [astro-ph.CO]} \BibitemShut {NoStop}%
\bibitem [{\citenamefont {Kobayashi}\ \emph {et~al.}(2022)\citenamefont {Kobayashi}, \citenamefont {Nishimichi}, \citenamefont {Takada},\ and\ \citenamefont {Miyatake}}]{Kobayashi:2021oud}%
  \BibitemOpen
  \bibfield  {author} {\bibinfo {author} {\bibfnamefont {Y.}~\bibnamefont {Kobayashi}}, \bibinfo {author} {\bibfnamefont {T.}~\bibnamefont {Nishimichi}}, \bibinfo {author} {\bibfnamefont {M.}~\bibnamefont {Takada}}, \ and\ \bibinfo {author} {\bibfnamefont {H.}~\bibnamefont {Miyatake}},\ }\href {\doibase 10.1103/PhysRevD.105.083517} {\bibfield  {journal} {\bibinfo  {journal} {Phys. Rev. D}\ }\textbf {\bibinfo {volume} {105}},\ \bibinfo {pages} {083517} (\bibinfo {year} {2022})},\ \Eprint {http://arxiv.org/abs/2110.06969} {arXiv:2110.06969 [astro-ph.CO]} \BibitemShut {NoStop}%
\bibitem [{\citenamefont {Cuesta-Lazaro}\ \emph {et~al.}(2023)\citenamefont {Cuesta-Lazaro} \emph {et~al.}}]{Cuesta-Lazaro:2023gbv}%
  \BibitemOpen
  \bibfield  {author} {\bibinfo {author} {\bibfnamefont {C.}~\bibnamefont {Cuesta-Lazaro}} \emph {et~al.},\ }\href@noop {} {\  (\bibinfo {year} {2023})},\ \Eprint {http://arxiv.org/abs/2309.16539} {arXiv:2309.16539 [astro-ph.CO]} \BibitemShut {NoStop}%
\bibitem [{\citenamefont {Valogiannis}\ \emph {et~al.}(2024)\citenamefont {Valogiannis}, \citenamefont {Yuan},\ and\ \citenamefont {Dvorkin}}]{Valogiannis:2023mxf}%
  \BibitemOpen
  \bibfield  {author} {\bibinfo {author} {\bibfnamefont {G.}~\bibnamefont {Valogiannis}}, \bibinfo {author} {\bibfnamefont {S.}~\bibnamefont {Yuan}}, \ and\ \bibinfo {author} {\bibfnamefont {C.}~\bibnamefont {Dvorkin}},\ }\href {\doibase 10.1103/PhysRevD.109.103503} {\bibfield  {journal} {\bibinfo  {journal} {Phys. Rev. D}\ }\textbf {\bibinfo {volume} {109}},\ \bibinfo {pages} {103503} (\bibinfo {year} {2024})},\ \Eprint {http://arxiv.org/abs/2310.16116} {arXiv:2310.16116 [astro-ph.CO]} \BibitemShut {NoStop}%
\bibitem [{\citenamefont {Hahn}\ \emph {et~al.}(2023)\citenamefont {Hahn}, \citenamefont {Eickenberg}, \citenamefont {Ho}, \citenamefont {Hou}, \citenamefont {Lemos}, \citenamefont {Massara}, \citenamefont {Modi}, \citenamefont {Moradinezhad~Dizgah}, \citenamefont {Parker},\ and\ \citenamefont {Blancard}}]{Hahn:2023kky}%
  \BibitemOpen
  \bibfield  {author} {\bibinfo {author} {\bibfnamefont {C.}~\bibnamefont {Hahn}}, \bibinfo {author} {\bibfnamefont {M.}~\bibnamefont {Eickenberg}}, \bibinfo {author} {\bibfnamefont {S.}~\bibnamefont {Ho}}, \bibinfo {author} {\bibfnamefont {J.}~\bibnamefont {Hou}}, \bibinfo {author} {\bibfnamefont {P.}~\bibnamefont {Lemos}}, \bibinfo {author} {\bibfnamefont {E.}~\bibnamefont {Massara}}, \bibinfo {author} {\bibfnamefont {C.}~\bibnamefont {Modi}}, \bibinfo {author} {\bibfnamefont {A.}~\bibnamefont {Moradinezhad~Dizgah}}, \bibinfo {author} {\bibfnamefont {L.}~\bibnamefont {Parker}}, \ and\ \bibinfo {author} {\bibfnamefont {B.~R.-S.}\ \bibnamefont {Blancard}},\ }\href@noop {} {\  (\bibinfo {year} {2023})},\ \Eprint {http://arxiv.org/abs/2310.15243} {arXiv:2310.15243 [astro-ph.CO]} \BibitemShut {NoStop}%
\bibitem [{\citenamefont {Hou}\ \emph {et~al.}(2024)\citenamefont {Hou}, \citenamefont {Moradinezhad~Dizgah}, \citenamefont {Hahn}, \citenamefont {Eickenberg}, \citenamefont {Ho}, \citenamefont {Lemos}, \citenamefont {Massara}, \citenamefont {Modi}, \citenamefont {Parker},\ and\ \citenamefont {Blancard}}]{Hou:2024blc}%
  \BibitemOpen
  \bibfield  {author} {\bibinfo {author} {\bibfnamefont {J.}~\bibnamefont {Hou}}, \bibinfo {author} {\bibfnamefont {A.}~\bibnamefont {Moradinezhad~Dizgah}}, \bibinfo {author} {\bibfnamefont {C.}~\bibnamefont {Hahn}}, \bibinfo {author} {\bibfnamefont {M.}~\bibnamefont {Eickenberg}}, \bibinfo {author} {\bibfnamefont {S.}~\bibnamefont {Ho}}, \bibinfo {author} {\bibfnamefont {P.}~\bibnamefont {Lemos}}, \bibinfo {author} {\bibfnamefont {E.}~\bibnamefont {Massara}}, \bibinfo {author} {\bibfnamefont {C.}~\bibnamefont {Modi}}, \bibinfo {author} {\bibfnamefont {L.}~\bibnamefont {Parker}}, \ and\ \bibinfo {author} {\bibfnamefont {B.~R.-S.}\ \bibnamefont {Blancard}},\ }\href {\doibase 10.1103/PhysRevD.109.103528} {\bibfield  {journal} {\bibinfo  {journal} {Phys. Rev. D}\ }\textbf {\bibinfo {volume} {109}},\ \bibinfo {pages} {103528} (\bibinfo {year} {2024})},\ \Eprint {http://arxiv.org/abs/2401.15074} {arXiv:2401.15074 [astro-ph.CO]} \BibitemShut {NoStop}%
\bibitem [{\citenamefont {Baumann}\ \emph {et~al.}(2012)\citenamefont {Baumann}, \citenamefont {Nicolis}, \citenamefont {Senatore},\ and\ \citenamefont {Zaldarriaga}}]{Baumann:2010tm}%
  \BibitemOpen
  \bibfield  {author} {\bibinfo {author} {\bibfnamefont {D.}~\bibnamefont {Baumann}}, \bibinfo {author} {\bibfnamefont {A.}~\bibnamefont {Nicolis}}, \bibinfo {author} {\bibfnamefont {L.}~\bibnamefont {Senatore}}, \ and\ \bibinfo {author} {\bibfnamefont {M.}~\bibnamefont {Zaldarriaga}},\ }\href {\doibase 10.1088/1475-7516/2012/07/051} {\bibfield  {journal} {\bibinfo  {journal} {JCAP}\ }\textbf {\bibinfo {volume} {1207}},\ \bibinfo {pages} {051} (\bibinfo {year} {2012})},\ \Eprint {http://arxiv.org/abs/1004.2488} {arXiv:1004.2488 [astro-ph.CO]} \BibitemShut {NoStop}%
\bibitem [{\citenamefont {Carrasco}\ \emph {et~al.}(2012)\citenamefont {Carrasco}, \citenamefont {Hertzberg},\ and\ \citenamefont {Senatore}}]{Carrasco:2012cv}%
  \BibitemOpen
  \bibfield  {author} {\bibinfo {author} {\bibfnamefont {J.~J.~M.}\ \bibnamefont {Carrasco}}, \bibinfo {author} {\bibfnamefont {M.~P.}\ \bibnamefont {Hertzberg}}, \ and\ \bibinfo {author} {\bibfnamefont {L.}~\bibnamefont {Senatore}},\ }\href {\doibase 10.1007/JHEP09(2012)082} {\bibfield  {journal} {\bibinfo  {journal} {JHEP}\ }\textbf {\bibinfo {volume} {09}},\ \bibinfo {pages} {082} (\bibinfo {year} {2012})},\ \Eprint {http://arxiv.org/abs/1206.2926} {arXiv:1206.2926 [astro-ph.CO]} \BibitemShut {NoStop}%
\bibitem [{\citenamefont {Ivanov}(2022)}]{Ivanov:2022mrd}%
  \BibitemOpen
  \bibfield  {author} {\bibinfo {author} {\bibfnamefont {M.~M.}\ \bibnamefont {Ivanov}},\ }\href@noop {} {\  (\bibinfo {year} {2022})},\ \Eprint {http://arxiv.org/abs/2212.08488} {arXiv:2212.08488 [astro-ph.CO]} \BibitemShut {NoStop}%
\bibitem [{\citenamefont {Desjacques}\ \emph {et~al.}(2018)\citenamefont {Desjacques}, \citenamefont {Jeong},\ and\ \citenamefont {Schmidt}}]{Desjacques:2016bnm}%
  \BibitemOpen
  \bibfield  {author} {\bibinfo {author} {\bibfnamefont {V.}~\bibnamefont {Desjacques}}, \bibinfo {author} {\bibfnamefont {D.}~\bibnamefont {Jeong}}, \ and\ \bibinfo {author} {\bibfnamefont {F.}~\bibnamefont {Schmidt}},\ }\href {\doibase 10.1016/j.physrep.2017.12.002} {\bibfield  {journal} {\bibinfo  {journal} {Phys. Rept.}\ }\textbf {\bibinfo {volume} {733}},\ \bibinfo {pages} {1} (\bibinfo {year} {2018})},\ \Eprint {http://arxiv.org/abs/1611.09787} {arXiv:1611.09787 [astro-ph.CO]} \BibitemShut {NoStop}%
\bibitem [{\citenamefont {Ivanov}\ \emph {et~al.}(2020{\natexlab{a}})\citenamefont {Ivanov}, \citenamefont {Simonovi\'c},\ and\ \citenamefont {Zaldarriaga}}]{Ivanov:2019pdj}%
  \BibitemOpen
  \bibfield  {author} {\bibinfo {author} {\bibfnamefont {M.~M.}\ \bibnamefont {Ivanov}}, \bibinfo {author} {\bibfnamefont {M.}~\bibnamefont {Simonovi\'c}}, \ and\ \bibinfo {author} {\bibfnamefont {M.}~\bibnamefont {Zaldarriaga}},\ }\href {\doibase 10.1088/1475-7516/2020/05/042} {\bibfield  {journal} {\bibinfo  {journal} {JCAP}\ }\textbf {\bibinfo {volume} {05}},\ \bibinfo {pages} {042} (\bibinfo {year} {2020}{\natexlab{a}})},\ \Eprint {http://arxiv.org/abs/1909.05277} {arXiv:1909.05277 [astro-ph.CO]} \BibitemShut {NoStop}%
\bibitem [{\citenamefont {D'Amico}\ \emph {et~al.}(2019)\citenamefont {D'Amico}, \citenamefont {Gleyzes}, \citenamefont {Kokron}, \citenamefont {Markovic}, \citenamefont {Senatore}, \citenamefont {Zhang}, \citenamefont {Beutler},\ and\ \citenamefont {Gil-Marín}}]{DAmico:2019fhj}%
  \BibitemOpen
  \bibfield  {author} {\bibinfo {author} {\bibfnamefont {G.}~\bibnamefont {D'Amico}}, \bibinfo {author} {\bibfnamefont {J.}~\bibnamefont {Gleyzes}}, \bibinfo {author} {\bibfnamefont {N.}~\bibnamefont {Kokron}}, \bibinfo {author} {\bibfnamefont {D.}~\bibnamefont {Markovic}}, \bibinfo {author} {\bibfnamefont {L.}~\bibnamefont {Senatore}}, \bibinfo {author} {\bibfnamefont {P.}~\bibnamefont {Zhang}}, \bibinfo {author} {\bibfnamefont {F.}~\bibnamefont {Beutler}}, \ and\ \bibinfo {author} {\bibfnamefont {H.}~\bibnamefont {Gil-Marín}},\ }\href@noop {} {\  (\bibinfo {year} {2019})},\ \Eprint {http://arxiv.org/abs/1909.05271} {arXiv:1909.05271 [astro-ph.CO]} \BibitemShut {NoStop}%
\bibitem [{\citenamefont {Chen}\ \emph {et~al.}(2022)\citenamefont {Chen}, \citenamefont {Vlah},\ and\ \citenamefont {White}}]{Chen:2021wdi}%
  \BibitemOpen
  \bibfield  {author} {\bibinfo {author} {\bibfnamefont {S.-F.}\ \bibnamefont {Chen}}, \bibinfo {author} {\bibfnamefont {Z.}~\bibnamefont {Vlah}}, \ and\ \bibinfo {author} {\bibfnamefont {M.}~\bibnamefont {White}},\ }\href {\doibase 10.1088/1475-7516/2022/02/008} {\bibfield  {journal} {\bibinfo  {journal} {JCAP}\ }\textbf {\bibinfo {volume} {02}},\ \bibinfo {pages} {008} (\bibinfo {year} {2022})},\ \Eprint {http://arxiv.org/abs/2110.05530} {arXiv:2110.05530 [astro-ph.CO]} \BibitemShut {NoStop}%
\bibitem [{\citenamefont {Philcox}\ and\ \citenamefont {Ivanov}(2022)}]{Philcox:2021kcw}%
  \BibitemOpen
  \bibfield  {author} {\bibinfo {author} {\bibfnamefont {O.~H.~E.}\ \bibnamefont {Philcox}}\ and\ \bibinfo {author} {\bibfnamefont {M.~M.}\ \bibnamefont {Ivanov}},\ }\href {\doibase 10.1103/PhysRevD.105.043517} {\bibfield  {journal} {\bibinfo  {journal} {Phys. Rev. D}\ }\textbf {\bibinfo {volume} {105}},\ \bibinfo {pages} {043517} (\bibinfo {year} {2022})},\ \Eprint {http://arxiv.org/abs/2112.04515} {arXiv:2112.04515 [astro-ph.CO]} \BibitemShut {NoStop}%
\bibitem [{\citenamefont {Ivanov}\ \emph {et~al.}(2023)\citenamefont {Ivanov}, \citenamefont {Philcox}, \citenamefont {Cabass}, \citenamefont {Nishimichi}, \citenamefont {Simonovi\'c},\ and\ \citenamefont {Zaldarriaga}}]{Ivanov:2023qzb}%
  \BibitemOpen
  \bibfield  {author} {\bibinfo {author} {\bibfnamefont {M.~M.}\ \bibnamefont {Ivanov}}, \bibinfo {author} {\bibfnamefont {O.~H.~E.}\ \bibnamefont {Philcox}}, \bibinfo {author} {\bibfnamefont {G.}~\bibnamefont {Cabass}}, \bibinfo {author} {\bibfnamefont {T.}~\bibnamefont {Nishimichi}}, \bibinfo {author} {\bibfnamefont {M.}~\bibnamefont {Simonovi\'c}}, \ and\ \bibinfo {author} {\bibfnamefont {M.}~\bibnamefont {Zaldarriaga}},\ }\href {\doibase 10.1103/PhysRevD.107.083515} {\bibfield  {journal} {\bibinfo  {journal} {Phys. Rev. D}\ }\textbf {\bibinfo {volume} {107}},\ \bibinfo {pages} {083515} (\bibinfo {year} {2023})},\ \Eprint {http://arxiv.org/abs/2302.04414} {arXiv:2302.04414 [astro-ph.CO]} \BibitemShut {NoStop}%
\bibitem [{\citenamefont {Chen}\ \emph {et~al.}(2024)\citenamefont {Chen}, \citenamefont {Ivanov}, \citenamefont {Philcox},\ and\ \citenamefont {Wenzl}}]{Chen:2024vuf}%
  \BibitemOpen
  \bibfield  {author} {\bibinfo {author} {\bibfnamefont {S.-F.}\ \bibnamefont {Chen}}, \bibinfo {author} {\bibfnamefont {M.~M.}\ \bibnamefont {Ivanov}}, \bibinfo {author} {\bibfnamefont {O.~H.~E.}\ \bibnamefont {Philcox}}, \ and\ \bibinfo {author} {\bibfnamefont {L.}~\bibnamefont {Wenzl}},\ }\href@noop {} {\  (\bibinfo {year} {2024})},\ \Eprint {http://arxiv.org/abs/2406.13388} {arXiv:2406.13388 [astro-ph.CO]} \BibitemShut {NoStop}%
\bibitem [{\citenamefont {Ivanov}\ \emph {et~al.}(2024{\natexlab{a}})\citenamefont {Ivanov}, \citenamefont {Cuesta-Lazaro}, \citenamefont {Mishra-Sharma}, \citenamefont {Obuljen},\ and\ \citenamefont {Toomey}}]{Ivanov:2024hgq}%
  \BibitemOpen
  \bibfield  {author} {\bibinfo {author} {\bibfnamefont {M.~M.}\ \bibnamefont {Ivanov}}, \bibinfo {author} {\bibfnamefont {C.}~\bibnamefont {Cuesta-Lazaro}}, \bibinfo {author} {\bibfnamefont {S.}~\bibnamefont {Mishra-Sharma}}, \bibinfo {author} {\bibfnamefont {A.}~\bibnamefont {Obuljen}}, \ and\ \bibinfo {author} {\bibfnamefont {M.~W.}\ \bibnamefont {Toomey}},\ }\href {\doibase 10.1103/PhysRevD.110.063538} {\bibfield  {journal} {\bibinfo  {journal} {Phys. Rev. D}\ }\textbf {\bibinfo {volume} {110}},\ \bibinfo {pages} {063538} (\bibinfo {year} {2024}{\natexlab{a}})},\ \Eprint {http://arxiv.org/abs/2402.13310} {arXiv:2402.13310 [astro-ph.CO]} \BibitemShut {NoStop}%
\bibitem [{\citenamefont {Ivanov}\ \emph {et~al.}(2025)\citenamefont {Ivanov}, \citenamefont {Obuljen}, \citenamefont {Cuesta-Lazaro},\ and\ \citenamefont {Toomey}}]{Ivanov:2024xgb}%
  \BibitemOpen
  \bibfield  {author} {\bibinfo {author} {\bibfnamefont {M.~M.}\ \bibnamefont {Ivanov}}, \bibinfo {author} {\bibfnamefont {A.}~\bibnamefont {Obuljen}}, \bibinfo {author} {\bibfnamefont {C.}~\bibnamefont {Cuesta-Lazaro}}, \ and\ \bibinfo {author} {\bibfnamefont {M.~W.}\ \bibnamefont {Toomey}},\ }\href {\doibase 10.1103/PhysRevD.111.063548} {\bibfield  {journal} {\bibinfo  {journal} {Phys. Rev. D}\ }\textbf {\bibinfo {volume} {111}},\ \bibinfo {pages} {063548} (\bibinfo {year} {2025})},\ \Eprint {http://arxiv.org/abs/2409.10609} {arXiv:2409.10609 [astro-ph.CO]} \BibitemShut {NoStop}%
\bibitem [{\citenamefont {Cabass}\ \emph {et~al.}(2024)\citenamefont {Cabass}, \citenamefont {Philcox}, \citenamefont {Ivanov}, \citenamefont {Akitsu}, \citenamefont {Chen}, \citenamefont {Simonovi\'c},\ and\ \citenamefont {Zaldarriaga}}]{Cabass:2024wob}%
  \BibitemOpen
  \bibfield  {author} {\bibinfo {author} {\bibfnamefont {G.}~\bibnamefont {Cabass}}, \bibinfo {author} {\bibfnamefont {O.~H.~E.}\ \bibnamefont {Philcox}}, \bibinfo {author} {\bibfnamefont {M.~M.}\ \bibnamefont {Ivanov}}, \bibinfo {author} {\bibfnamefont {K.}~\bibnamefont {Akitsu}}, \bibinfo {author} {\bibfnamefont {S.-F.}\ \bibnamefont {Chen}}, \bibinfo {author} {\bibfnamefont {M.}~\bibnamefont {Simonovi\'c}}, \ and\ \bibinfo {author} {\bibfnamefont {M.}~\bibnamefont {Zaldarriaga}},\ }\href@noop {} {\  (\bibinfo {year} {2024})},\ \Eprint {http://arxiv.org/abs/2404.01894} {arXiv:2404.01894 [astro-ph.CO]} \BibitemShut {NoStop}%
\bibitem [{\citenamefont {Ivanov}\ \emph {et~al.}(2024{\natexlab{b}})\citenamefont {Ivanov} \emph {et~al.}}]{Ivanov:2024dgv}%
  \BibitemOpen
  \bibfield  {author} {\bibinfo {author} {\bibfnamefont {M.~M.}\ \bibnamefont {Ivanov}} \emph {et~al.},\ }\href@noop {} {\  (\bibinfo {year} {2024}{\natexlab{b}})},\ \Eprint {http://arxiv.org/abs/2412.01888} {arXiv:2412.01888 [astro-ph.CO]} \BibitemShut {NoStop}%
\bibitem [{\citenamefont {Ivanov}(2025)}]{Ivanov:2025qie}%
  \BibitemOpen
  \bibfield  {author} {\bibinfo {author} {\bibfnamefont {M.~M.}\ \bibnamefont {Ivanov}},\ }\href@noop {} {\  (\bibinfo {year} {2025})},\ \Eprint {http://arxiv.org/abs/2503.07270} {arXiv:2503.07270 [astro-ph.CO]} \BibitemShut {NoStop}%
\bibitem [{\citenamefont {Sullivan}\ \emph {et~al.}(2025)\citenamefont {Sullivan}, \citenamefont {Cuesta-Lazaro}, \citenamefont {Ivanov}, \citenamefont {Ni}, \citenamefont {Bose}, \citenamefont {Hadzhiyska}, \citenamefont {Hern{\'a}ndez-Aguayo}, \citenamefont {Hernquist},\ and\ \citenamefont {Kannan}}]{Sullivan:2025eei}%
  \BibitemOpen
  \bibfield  {author} {\bibinfo {author} {\bibfnamefont {J.~M.}\ \bibnamefont {Sullivan}}, \bibinfo {author} {\bibfnamefont {C.}~\bibnamefont {Cuesta-Lazaro}}, \bibinfo {author} {\bibfnamefont {M.~M.}\ \bibnamefont {Ivanov}}, \bibinfo {author} {\bibfnamefont {Y.}~\bibnamefont {Ni}}, \bibinfo {author} {\bibfnamefont {S.}~\bibnamefont {Bose}}, \bibinfo {author} {\bibfnamefont {B.}~\bibnamefont {Hadzhiyska}}, \bibinfo {author} {\bibfnamefont {C.}~\bibnamefont {Hern{\'a}ndez-Aguayo}}, \bibinfo {author} {\bibfnamefont {L.}~\bibnamefont {Hernquist}}, \ and\ \bibinfo {author} {\bibfnamefont {R.}~\bibnamefont {Kannan}},\ }\href@noop {} {\  (\bibinfo {year} {2025})},\ \Eprint {http://arxiv.org/abs/2505.03626} {arXiv:2505.03626 [astro-ph.CO]} \BibitemShut {NoStop}%
\bibitem [{\citenamefont {Sullivan}\ \emph {et~al.}(2021)\citenamefont {Sullivan}, \citenamefont {Seljak},\ and\ \citenamefont {Singh}}]{Sullivan:2021sof}%
  \BibitemOpen
  \bibfield  {author} {\bibinfo {author} {\bibfnamefont {J.~M.}\ \bibnamefont {Sullivan}}, \bibinfo {author} {\bibfnamefont {U.}~\bibnamefont {Seljak}}, \ and\ \bibinfo {author} {\bibfnamefont {S.}~\bibnamefont {Singh}},\ }\href {\doibase 10.1088/1475-7516/2021/11/026} {\bibfield  {journal} {\bibinfo  {journal} {JCAP}\ }\textbf {\bibinfo {volume} {11}},\ \bibinfo {pages} {026} (\bibinfo {year} {2021})},\ \Eprint {http://arxiv.org/abs/2104.10676} {arXiv:2104.10676 [astro-ph.CO]} \BibitemShut {NoStop}%
\bibitem [{\citenamefont {Obuljen}\ \emph {et~al.}(2023)\citenamefont {Obuljen}, \citenamefont {Simonovi\'c}, \citenamefont {Schneider},\ and\ \citenamefont {Feldmann}}]{Obuljen:2022cjo}%
  \BibitemOpen
  \bibfield  {author} {\bibinfo {author} {\bibfnamefont {A.}~\bibnamefont {Obuljen}}, \bibinfo {author} {\bibfnamefont {M.}~\bibnamefont {Simonovi\'c}}, \bibinfo {author} {\bibfnamefont {A.}~\bibnamefont {Schneider}}, \ and\ \bibinfo {author} {\bibfnamefont {R.}~\bibnamefont {Feldmann}},\ }\href {\doibase 10.1103/PhysRevD.108.083528} {\bibfield  {journal} {\bibinfo  {journal} {Phys. Rev. D}\ }\textbf {\bibinfo {volume} {108}},\ \bibinfo {pages} {083528} (\bibinfo {year} {2023})},\ \Eprint {http://arxiv.org/abs/2207.12398} {arXiv:2207.12398 [astro-ph.CO]} \BibitemShut {NoStop}%
\bibitem [{\citenamefont {Modi}\ and\ \citenamefont {Philcox}(2023)}]{Modi:2023drt}%
  \BibitemOpen
  \bibfield  {author} {\bibinfo {author} {\bibfnamefont {C.}~\bibnamefont {Modi}}\ and\ \bibinfo {author} {\bibfnamefont {O.~H.~E.}\ \bibnamefont {Philcox}},\ }\href@noop {} {\  (\bibinfo {year} {2023})},\ \Eprint {http://arxiv.org/abs/2309.10270} {arXiv:2309.10270 [astro-ph.CO]} \BibitemShut {NoStop}%
\bibitem [{\citenamefont {Akitsu}(2024)}]{Akitsu:2024lyt}%
  \BibitemOpen
  \bibfield  {author} {\bibinfo {author} {\bibfnamefont {K.}~\bibnamefont {Akitsu}},\ }\href@noop {} {\  (\bibinfo {year} {2024})},\ \Eprint {http://arxiv.org/abs/2410.08998} {arXiv:2410.08998 [astro-ph.CO]} \BibitemShut {NoStop}%
\bibitem [{\citenamefont {Zhang}\ \emph {et~al.}(2025{\natexlab{a}})\citenamefont {Zhang}, \citenamefont {Bonici}, \citenamefont {D'Amico}, \citenamefont {Paradiso},\ and\ \citenamefont {Percival}}]{Zhang:2024thl}%
  \BibitemOpen
  \bibfield  {author} {\bibinfo {author} {\bibfnamefont {H.}~\bibnamefont {Zhang}}, \bibinfo {author} {\bibfnamefont {M.}~\bibnamefont {Bonici}}, \bibinfo {author} {\bibfnamefont {G.}~\bibnamefont {D'Amico}}, \bibinfo {author} {\bibfnamefont {S.}~\bibnamefont {Paradiso}}, \ and\ \bibinfo {author} {\bibfnamefont {W.~J.}\ \bibnamefont {Percival}},\ }\href {\doibase 10.1088/1475-7516/2025/04/041} {\bibfield  {journal} {\bibinfo  {journal} {JCAP}\ }\textbf {\bibinfo {volume} {04}},\ \bibinfo {pages} {041} (\bibinfo {year} {2025}{\natexlab{a}})},\ \Eprint {http://arxiv.org/abs/2409.12937} {arXiv:2409.12937 [astro-ph.CO]} \BibitemShut {NoStop}%
\bibitem [{\citenamefont {Zhang}\ \emph {et~al.}(2025{\natexlab{b}})\citenamefont {Zhang} \emph {et~al.}}]{DESI:2025wzd}%
  \BibitemOpen
  \bibfield  {author} {\bibinfo {author} {\bibfnamefont {H.}~\bibnamefont {Zhang}} \emph {et~al.} (\bibinfo {collaboration} {DESI}),\ }\href@noop {} {\  (\bibinfo {year} {2025}{\natexlab{b}})},\ \Eprint {http://arxiv.org/abs/2504.10407} {arXiv:2504.10407 [astro-ph.CO]} \BibitemShut {NoStop}%
\bibitem [{\citenamefont {Zhang}\ \emph {et~al.}(2025{\natexlab{c}})\citenamefont {Zhang}, \citenamefont {Modi},\ and\ \citenamefont {Philcox}}]{Zhang:2025sfk}%
  \BibitemOpen
  \bibfield  {author} {\bibinfo {author} {\bibfnamefont {G.}~\bibnamefont {Zhang}}, \bibinfo {author} {\bibfnamefont {C.}~\bibnamefont {Modi}}, \ and\ \bibinfo {author} {\bibfnamefont {O.~H.~E.}\ \bibnamefont {Philcox}},\ }\href@noop {} {\  (\bibinfo {year} {2025}{\natexlab{c}})},\ \Eprint {http://arxiv.org/abs/2505.13591} {arXiv:2505.13591 [astro-ph.CO]} \BibitemShut {NoStop}%
\bibitem [{\citenamefont {Schmittfull}\ \emph {et~al.}(2019)\citenamefont {Schmittfull}, \citenamefont {Simonović}, \citenamefont {Assassi},\ and\ \citenamefont {Zaldarriaga}}]{Schmittfull:2018yuk}%
  \BibitemOpen
  \bibfield  {author} {\bibinfo {author} {\bibfnamefont {M.}~\bibnamefont {Schmittfull}}, \bibinfo {author} {\bibfnamefont {M.}~\bibnamefont {Simonović}}, \bibinfo {author} {\bibfnamefont {V.}~\bibnamefont {Assassi}}, \ and\ \bibinfo {author} {\bibfnamefont {M.}~\bibnamefont {Zaldarriaga}},\ }\href {\doibase 10.1103/PhysRevD.100.043514} {\bibfield  {journal} {\bibinfo  {journal} {Phys.\ Rev.\ D}\ }\textbf {\bibinfo {volume} {100}},\ \bibinfo {pages} {043514} (\bibinfo {year} {2019})},\ \Eprint {http://arxiv.org/abs/1811.10640} {arXiv:1811.10640 [astro-ph.CO]} \BibitemShut {NoStop}%
\bibitem [{\citenamefont {Schmittfull}\ \emph {et~al.}(2021)\citenamefont {Schmittfull}, \citenamefont {Simonovi\'c}, \citenamefont {Ivanov}, \citenamefont {Philcox},\ and\ \citenamefont {Zaldarriaga}}]{Schmittfull:2020trd}%
  \BibitemOpen
  \bibfield  {author} {\bibinfo {author} {\bibfnamefont {M.}~\bibnamefont {Schmittfull}}, \bibinfo {author} {\bibfnamefont {M.}~\bibnamefont {Simonovi\'c}}, \bibinfo {author} {\bibfnamefont {M.~M.}\ \bibnamefont {Ivanov}}, \bibinfo {author} {\bibfnamefont {O.~H.~E.}\ \bibnamefont {Philcox}}, \ and\ \bibinfo {author} {\bibfnamefont {M.}~\bibnamefont {Zaldarriaga}},\ }\href {\doibase 10.1088/1475-7516/2021/05/059} {\bibfield  {journal} {\bibinfo  {journal} {JCAP}\ }\textbf {\bibinfo {volume} {05}},\ \bibinfo {pages} {059} (\bibinfo {year} {2021})},\ \Eprint {http://arxiv.org/abs/2012.03334} {arXiv:2012.03334 [astro-ph.CO]} \BibitemShut {NoStop}%
\bibitem [{\citenamefont {Foreman}\ \emph {et~al.}(2024)\citenamefont {Foreman}, \citenamefont {Obuljen},\ and\ \citenamefont {Simonovi\'c}}]{Foreman:2024kzw}%
  \BibitemOpen
  \bibfield  {author} {\bibinfo {author} {\bibfnamefont {S.}~\bibnamefont {Foreman}}, \bibinfo {author} {\bibfnamefont {A.}~\bibnamefont {Obuljen}}, \ and\ \bibinfo {author} {\bibfnamefont {M.}~\bibnamefont {Simonovi\'c}},\ }\href@noop {} {\  (\bibinfo {year} {2024})},\ \Eprint {http://arxiv.org/abs/2405.18559} {arXiv:2405.18559 [astro-ph.CO]} \BibitemShut {NoStop}%
\bibitem [{\citenamefont {Schmittfull}\ \emph {et~al.}(2015)\citenamefont {Schmittfull}, \citenamefont {Baldauf},\ and\ \citenamefont {Seljak}}]{Schmittfull:2014tca}%
  \BibitemOpen
  \bibfield  {author} {\bibinfo {author} {\bibfnamefont {M.}~\bibnamefont {Schmittfull}}, \bibinfo {author} {\bibfnamefont {T.}~\bibnamefont {Baldauf}}, \ and\ \bibinfo {author} {\bibfnamefont {U.}~\bibnamefont {Seljak}},\ }\href {\doibase 10.1103/PhysRevD.91.043530} {\bibfield  {journal} {\bibinfo  {journal} {Phys. Rev. D}\ }\textbf {\bibinfo {volume} {91}},\ \bibinfo {pages} {043530} (\bibinfo {year} {2015})},\ \Eprint {http://arxiv.org/abs/1411.6595} {arXiv:1411.6595 [astro-ph.CO]} \BibitemShut {NoStop}%
\bibitem [{\citenamefont {Lazeyras}\ and\ \citenamefont {Schmidt}(2018)}]{Lazeyras:2017hxw}%
  \BibitemOpen
  \bibfield  {author} {\bibinfo {author} {\bibfnamefont {T.}~\bibnamefont {Lazeyras}}\ and\ \bibinfo {author} {\bibfnamefont {F.}~\bibnamefont {Schmidt}},\ }\href {\doibase 10.1088/1475-7516/2018/09/008} {\bibfield  {journal} {\bibinfo  {journal} {JCAP}\ }\textbf {\bibinfo {volume} {1809}},\ \bibinfo {pages} {008} (\bibinfo {year} {2018})},\ \Eprint {http://arxiv.org/abs/1712.07531} {arXiv:1712.07531 [astro-ph.CO]} \BibitemShut {NoStop}%
\bibitem [{\citenamefont {Abidi}\ and\ \citenamefont {Baldauf}(2018)}]{Abidi:2018eyd}%
  \BibitemOpen
  \bibfield  {author} {\bibinfo {author} {\bibfnamefont {M.~M.}\ \bibnamefont {Abidi}}\ and\ \bibinfo {author} {\bibfnamefont {T.}~\bibnamefont {Baldauf}},\ }\href {\doibase 10.1088/1475-7516/2018/07/029} {\bibfield  {journal} {\bibinfo  {journal} {JCAP}\ }\textbf {\bibinfo {volume} {1807}},\ \bibinfo {pages} {029} (\bibinfo {year} {2018})},\ \Eprint {http://arxiv.org/abs/1802.07622} {arXiv:1802.07622 [astro-ph.CO]} \BibitemShut {NoStop}%
\bibitem [{\citenamefont {Schmidt}\ \emph {et~al.}(2019)\citenamefont {Schmidt}, \citenamefont {Elsner}, \citenamefont {Jasche}, \citenamefont {Nguyen},\ and\ \citenamefont {Lavaux}}]{Schmidt:2018bkr}%
  \BibitemOpen
  \bibfield  {author} {\bibinfo {author} {\bibfnamefont {F.}~\bibnamefont {Schmidt}}, \bibinfo {author} {\bibfnamefont {F.}~\bibnamefont {Elsner}}, \bibinfo {author} {\bibfnamefont {J.}~\bibnamefont {Jasche}}, \bibinfo {author} {\bibfnamefont {N.~M.}\ \bibnamefont {Nguyen}}, \ and\ \bibinfo {author} {\bibfnamefont {G.}~\bibnamefont {Lavaux}},\ }\href {\doibase 10.1088/1475-7516/2019/01/042} {\bibfield  {journal} {\bibinfo  {journal} {JCAP}\ }\textbf {\bibinfo {volume} {01}},\ \bibinfo {pages} {042} (\bibinfo {year} {2019})},\ \Eprint {http://arxiv.org/abs/1808.02002} {arXiv:1808.02002 [astro-ph.CO]} \BibitemShut {NoStop}%
\bibitem [{\citenamefont {Elsner}\ \emph {et~al.}(2020)\citenamefont {Elsner}, \citenamefont {Schmidt}, \citenamefont {Jasche}, \citenamefont {Lavaux},\ and\ \citenamefont {Nguyen}}]{Elsner:2019rql}%
  \BibitemOpen
  \bibfield  {author} {\bibinfo {author} {\bibfnamefont {F.}~\bibnamefont {Elsner}}, \bibinfo {author} {\bibfnamefont {F.}~\bibnamefont {Schmidt}}, \bibinfo {author} {\bibfnamefont {J.}~\bibnamefont {Jasche}}, \bibinfo {author} {\bibfnamefont {G.}~\bibnamefont {Lavaux}}, \ and\ \bibinfo {author} {\bibfnamefont {N.-M.}\ \bibnamefont {Nguyen}},\ }\href {\doibase 10.1088/1475-7516/2020/01/029} {\bibfield  {journal} {\bibinfo  {journal} {JCAP}\ }\textbf {\bibinfo {volume} {01}},\ \bibinfo {pages} {029} (\bibinfo {year} {2020})},\ \Eprint {http://arxiv.org/abs/1906.07143} {arXiv:1906.07143 [astro-ph.CO]} \BibitemShut {NoStop}%
\bibitem [{\citenamefont {Cabass}\ and\ \citenamefont {Schmidt}(2020)}]{Cabass:2019lqx}%
  \BibitemOpen
  \bibfield  {author} {\bibinfo {author} {\bibfnamefont {G.}~\bibnamefont {Cabass}}\ and\ \bibinfo {author} {\bibfnamefont {F.}~\bibnamefont {Schmidt}},\ }\href {\doibase 10.1088/1475-7516/2020/04/042} {\bibfield  {journal} {\bibinfo  {journal} {JCAP}\ }\textbf {\bibinfo {volume} {04}},\ \bibinfo {pages} {042} (\bibinfo {year} {2020})},\ \Eprint {http://arxiv.org/abs/1909.04022} {arXiv:1909.04022 [astro-ph.CO]} \BibitemShut {NoStop}%
\bibitem [{\citenamefont {Modi}\ \emph {et~al.}(2020)\citenamefont {Modi}, \citenamefont {Chen},\ and\ \citenamefont {White}}]{Modi:2019qbt}%
  \BibitemOpen
  \bibfield  {author} {\bibinfo {author} {\bibfnamefont {C.}~\bibnamefont {Modi}}, \bibinfo {author} {\bibfnamefont {S.-F.}\ \bibnamefont {Chen}}, \ and\ \bibinfo {author} {\bibfnamefont {M.}~\bibnamefont {White}},\ }\href {\doibase 10.1093/mnras/staa251} {\bibfield  {journal} {\bibinfo  {journal} {Mon. Not. Roy. Astron. Soc.}\ }\textbf {\bibinfo {volume} {492}},\ \bibinfo {pages} {5754} (\bibinfo {year} {2020})},\ \Eprint {http://arxiv.org/abs/1910.07097} {arXiv:1910.07097 [astro-ph.CO]} \BibitemShut {NoStop}%
\bibitem [{\citenamefont {Schmidt}(2020)}]{Schmidt:2020tao}%
  \BibitemOpen
  \bibfield  {author} {\bibinfo {author} {\bibfnamefont {F.}~\bibnamefont {Schmidt}},\ }\href@noop {} {\  (\bibinfo {year} {2020})},\ \Eprint {http://arxiv.org/abs/2009.14176} {arXiv:2009.14176 [astro-ph.CO]} \BibitemShut {NoStop}%
\bibitem [{\citenamefont {Schmidt}\ \emph {et~al.}(2020)\citenamefont {Schmidt}, \citenamefont {Cabass}, \citenamefont {Jasche},\ and\ \citenamefont {Lavaux}}]{Schmidt:2020viy}%
  \BibitemOpen
  \bibfield  {author} {\bibinfo {author} {\bibfnamefont {F.}~\bibnamefont {Schmidt}}, \bibinfo {author} {\bibfnamefont {G.}~\bibnamefont {Cabass}}, \bibinfo {author} {\bibfnamefont {J.}~\bibnamefont {Jasche}}, \ and\ \bibinfo {author} {\bibfnamefont {G.}~\bibnamefont {Lavaux}},\ }\href {\doibase 10.1088/1475-7516/2020/11/008} {\bibfield  {journal} {\bibinfo  {journal} {JCAP}\ }\textbf {\bibinfo {volume} {11}},\ \bibinfo {pages} {008} (\bibinfo {year} {2020})},\ \Eprint {http://arxiv.org/abs/2004.06707} {arXiv:2004.06707 [astro-ph.CO]} \BibitemShut {NoStop}%
\bibitem [{\citenamefont {Lazeyras}\ \emph {et~al.}(2021)\citenamefont {Lazeyras}, \citenamefont {Barreira},\ and\ \citenamefont {Schmidt}}]{Lazeyras:2021dar}%
  \BibitemOpen
  \bibfield  {author} {\bibinfo {author} {\bibfnamefont {T.}~\bibnamefont {Lazeyras}}, \bibinfo {author} {\bibfnamefont {A.}~\bibnamefont {Barreira}}, \ and\ \bibinfo {author} {\bibfnamefont {F.}~\bibnamefont {Schmidt}},\ }\href {\doibase 10.1088/1475-7516/2021/10/063} {\bibfield  {journal} {\bibinfo  {journal} {JCAP}\ }\textbf {\bibinfo {volume} {10}},\ \bibinfo {pages} {063} (\bibinfo {year} {2021})},\ \Eprint {http://arxiv.org/abs/2106.14713} {arXiv:2106.14713 [astro-ph.CO]} \BibitemShut {NoStop}%
\bibitem [{\citenamefont {Stadler}\ \emph {et~al.}(2023)\citenamefont {Stadler}, \citenamefont {Schmidt},\ and\ \citenamefont {Reinecke}}]{Stadler:2023hea}%
  \BibitemOpen
  \bibfield  {author} {\bibinfo {author} {\bibfnamefont {J.}~\bibnamefont {Stadler}}, \bibinfo {author} {\bibfnamefont {F.}~\bibnamefont {Schmidt}}, \ and\ \bibinfo {author} {\bibfnamefont {M.}~\bibnamefont {Reinecke}},\ }\href {\doibase 10.1088/1475-7516/2023/10/069} {\bibfield  {journal} {\bibinfo  {journal} {JCAP}\ }\textbf {\bibinfo {volume} {10}},\ \bibinfo {pages} {069} (\bibinfo {year} {2023})},\ \Eprint {http://arxiv.org/abs/2303.09876} {arXiv:2303.09876 [astro-ph.CO]} \BibitemShut {NoStop}%
\bibitem [{\citenamefont {Nguyen}\ \emph {et~al.}(2024)\citenamefont {Nguyen}, \citenamefont {Schmidt}, \citenamefont {Tucci}, \citenamefont {Reinecke},\ and\ \citenamefont {Kosti\'c}}]{Nguyen:2024yth}%
  \BibitemOpen
  \bibfield  {author} {\bibinfo {author} {\bibfnamefont {N.-M.}\ \bibnamefont {Nguyen}}, \bibinfo {author} {\bibfnamefont {F.}~\bibnamefont {Schmidt}}, \bibinfo {author} {\bibfnamefont {B.}~\bibnamefont {Tucci}}, \bibinfo {author} {\bibfnamefont {M.}~\bibnamefont {Reinecke}}, \ and\ \bibinfo {author} {\bibfnamefont {A.}~\bibnamefont {Kosti\'c}},\ }\href@noop {} {\  (\bibinfo {year} {2024})},\ \Eprint {http://arxiv.org/abs/2403.03220} {arXiv:2403.03220 [astro-ph.CO]} \BibitemShut {NoStop}%
\bibitem [{\citenamefont {Chudaykin}\ \emph {et~al.}(2021{\natexlab{a}})\citenamefont {Chudaykin}, \citenamefont {Dolgikh},\ and\ \citenamefont {Ivanov}}]{Chudaykin:2020ghx}%
  \BibitemOpen
  \bibfield  {author} {\bibinfo {author} {\bibfnamefont {A.}~\bibnamefont {Chudaykin}}, \bibinfo {author} {\bibfnamefont {K.}~\bibnamefont {Dolgikh}}, \ and\ \bibinfo {author} {\bibfnamefont {M.~M.}\ \bibnamefont {Ivanov}},\ }\href {\doibase 10.1103/PhysRevD.103.023507} {\bibfield  {journal} {\bibinfo  {journal} {Phys. Rev. D}\ }\textbf {\bibinfo {volume} {103}},\ \bibinfo {pages} {023507} (\bibinfo {year} {2021}{\natexlab{a}})},\ \Eprint {http://arxiv.org/abs/2009.10106} {arXiv:2009.10106 [astro-ph.CO]} \BibitemShut {NoStop}%
\bibitem [{\citenamefont {Philcox}\ \emph {et~al.}(2021)\citenamefont {Philcox}, \citenamefont {Ivanov}, \citenamefont {Zaldarriaga}, \citenamefont {Simonovic},\ and\ \citenamefont {Schmittfull}}]{Philcox:2020zyp}%
  \BibitemOpen
  \bibfield  {author} {\bibinfo {author} {\bibfnamefont {O.~H.~E.}\ \bibnamefont {Philcox}}, \bibinfo {author} {\bibfnamefont {M.~M.}\ \bibnamefont {Ivanov}}, \bibinfo {author} {\bibfnamefont {M.}~\bibnamefont {Zaldarriaga}}, \bibinfo {author} {\bibfnamefont {M.}~\bibnamefont {Simonovic}}, \ and\ \bibinfo {author} {\bibfnamefont {M.}~\bibnamefont {Schmittfull}},\ }\href {\doibase 10.1103/PhysRevD.103.043508} {\bibfield  {journal} {\bibinfo  {journal} {Phys. Rev. D}\ }\textbf {\bibinfo {volume} {103}},\ \bibinfo {pages} {043508} (\bibinfo {year} {2021})},\ \Eprint {http://arxiv.org/abs/2009.03311} {arXiv:2009.03311 [astro-ph.CO]} \BibitemShut {NoStop}%
\bibitem [{\citenamefont {Aghanim}\ \emph {et~al.}(2020{\natexlab{a}})\citenamefont {Aghanim} \emph {et~al.}}]{Planck:2018vyg}%
  \BibitemOpen
  \bibfield  {author} {\bibinfo {author} {\bibfnamefont {N.}~\bibnamefont {Aghanim}} \emph {et~al.} (\bibinfo {collaboration} {Planck}),\ }\href {\doibase 10.1051/0004-6361/201833910} {\bibfield  {journal} {\bibinfo  {journal} {Astron. Astrophys.}\ }\textbf {\bibinfo {volume} {641}},\ \bibinfo {pages} {A6} (\bibinfo {year} {2020}{\natexlab{a}})},\ \bibinfo {note} {[Erratum: Astron.Astrophys. 652, C4 (2021)]},\ \Eprint {http://arxiv.org/abs/1807.06209} {arXiv:1807.06209 [astro-ph.CO]} \BibitemShut {NoStop}%
\bibitem [{\citenamefont {Scolnic}\ \emph {et~al.}(2022)\citenamefont {Scolnic} \emph {et~al.}}]{Scolnic:2021amr}%
  \BibitemOpen
  \bibfield  {author} {\bibinfo {author} {\bibfnamefont {D.}~\bibnamefont {Scolnic}} \emph {et~al.},\ }\href {\doibase 10.3847/1538-4357/ac8b7a} {\bibfield  {journal} {\bibinfo  {journal} {Astrophys. J.}\ }\textbf {\bibinfo {volume} {938}},\ \bibinfo {pages} {113} (\bibinfo {year} {2022})},\ \Eprint {http://arxiv.org/abs/2112.03863} {arXiv:2112.03863 [astro-ph.CO]} \BibitemShut {NoStop}%
\bibitem [{\citenamefont {Brout}\ \emph {et~al.}(2022)\citenamefont {Brout} \emph {et~al.}}]{Brout:2022vxf}%
  \BibitemOpen
  \bibfield  {author} {\bibinfo {author} {\bibfnamefont {D.}~\bibnamefont {Brout}} \emph {et~al.},\ }\href {\doibase 10.3847/1538-4357/ac8e04} {\bibfield  {journal} {\bibinfo  {journal} {Astrophys. J.}\ }\textbf {\bibinfo {volume} {938}},\ \bibinfo {pages} {110} (\bibinfo {year} {2022})},\ \Eprint {http://arxiv.org/abs/2202.04077} {arXiv:2202.04077 [astro-ph.CO]} \BibitemShut {NoStop}%
\bibitem [{\citenamefont {Ivanov}\ \emph {et~al.}(2020{\natexlab{b}})\citenamefont {Ivanov}, \citenamefont {Simonovi\'c},\ and\ \citenamefont {Zaldarriaga}}]{Ivanov:2019hqk}%
  \BibitemOpen
  \bibfield  {author} {\bibinfo {author} {\bibfnamefont {M.~M.}\ \bibnamefont {Ivanov}}, \bibinfo {author} {\bibfnamefont {M.}~\bibnamefont {Simonovi\'c}}, \ and\ \bibinfo {author} {\bibfnamefont {M.}~\bibnamefont {Zaldarriaga}},\ }\href {\doibase 10.1103/PhysRevD.101.083504} {\bibfield  {journal} {\bibinfo  {journal} {Phys. Rev. D}\ }\textbf {\bibinfo {volume} {101}},\ \bibinfo {pages} {083504} (\bibinfo {year} {2020}{\natexlab{b}})},\ \Eprint {http://arxiv.org/abs/1912.08208} {arXiv:1912.08208 [astro-ph.CO]} \BibitemShut {NoStop}%
\bibitem [{\citenamefont {Chudaykin}\ \emph {et~al.}(2021{\natexlab{b}})\citenamefont {Chudaykin}, \citenamefont {Ivanov},\ and\ \citenamefont {Simonovi\'c}}]{Chudaykin:2020hbf}%
  \BibitemOpen
  \bibfield  {author} {\bibinfo {author} {\bibfnamefont {A.}~\bibnamefont {Chudaykin}}, \bibinfo {author} {\bibfnamefont {M.~M.}\ \bibnamefont {Ivanov}}, \ and\ \bibinfo {author} {\bibfnamefont {M.}~\bibnamefont {Simonovi\'c}},\ }\href {\doibase 10.1103/PhysRevD.103.043525} {\bibfield  {journal} {\bibinfo  {journal} {Phys. Rev. D}\ }\textbf {\bibinfo {volume} {103}},\ \bibinfo {pages} {043525} (\bibinfo {year} {2021}{\natexlab{b}})},\ \Eprint {http://arxiv.org/abs/2009.10724} {arXiv:2009.10724 [astro-ph.CO]} \BibitemShut {NoStop}%
\bibitem [{\citenamefont {Ivanov}\ \emph {et~al.}(2022{\natexlab{a}})\citenamefont {Ivanov}, \citenamefont {Philcox}, \citenamefont {Nishimichi}, \citenamefont {Simonovi\'c}, \citenamefont {Takada},\ and\ \citenamefont {Zaldarriaga}}]{Ivanov:2021kcd}%
  \BibitemOpen
  \bibfield  {author} {\bibinfo {author} {\bibfnamefont {M.~M.}\ \bibnamefont {Ivanov}}, \bibinfo {author} {\bibfnamefont {O.~H.~E.}\ \bibnamefont {Philcox}}, \bibinfo {author} {\bibfnamefont {T.}~\bibnamefont {Nishimichi}}, \bibinfo {author} {\bibfnamefont {M.}~\bibnamefont {Simonovi\'c}}, \bibinfo {author} {\bibfnamefont {M.}~\bibnamefont {Takada}}, \ and\ \bibinfo {author} {\bibfnamefont {M.}~\bibnamefont {Zaldarriaga}},\ }\href {\doibase 10.1103/PhysRevD.105.063512} {\bibfield  {journal} {\bibinfo  {journal} {Phys. Rev. D}\ }\textbf {\bibinfo {volume} {105}},\ \bibinfo {pages} {063512} (\bibinfo {year} {2022}{\natexlab{a}})},\ \Eprint {http://arxiv.org/abs/2110.10161} {arXiv:2110.10161 [astro-ph.CO]} \BibitemShut {NoStop}%
\bibitem [{\citenamefont {Bernardeau}\ \emph {et~al.}(2002)\citenamefont {Bernardeau}, \citenamefont {Colombi}, \citenamefont {Gaztanaga},\ and\ \citenamefont {Scoccimarro}}]{Bernardeau:2001qr}%
  \BibitemOpen
  \bibfield  {author} {\bibinfo {author} {\bibfnamefont {F.}~\bibnamefont {Bernardeau}}, \bibinfo {author} {\bibfnamefont {S.}~\bibnamefont {Colombi}}, \bibinfo {author} {\bibfnamefont {E.}~\bibnamefont {Gaztanaga}}, \ and\ \bibinfo {author} {\bibfnamefont {R.}~\bibnamefont {Scoccimarro}},\ }\href {\doibase 10.1016/S0370-1573(02)00135-7} {\bibfield  {journal} {\bibinfo  {journal} {Phys. Rept.}\ }\textbf {\bibinfo {volume} {367}},\ \bibinfo {pages} {1} (\bibinfo {year} {2002})},\ \Eprint {http://arxiv.org/abs/astro-ph/0112551} {arXiv:astro-ph/0112551 [astro-ph]} \BibitemShut {NoStop}%
\bibitem [{\citenamefont {Chudaykin}\ \emph {et~al.}(2020)\citenamefont {Chudaykin}, \citenamefont {Ivanov}, \citenamefont {Philcox},\ and\ \citenamefont {Simonovi\'c}}]{Chudaykin:2020aoj}%
  \BibitemOpen
  \bibfield  {author} {\bibinfo {author} {\bibfnamefont {A.}~\bibnamefont {Chudaykin}}, \bibinfo {author} {\bibfnamefont {M.~M.}\ \bibnamefont {Ivanov}}, \bibinfo {author} {\bibfnamefont {O.~H.~E.}\ \bibnamefont {Philcox}}, \ and\ \bibinfo {author} {\bibfnamefont {M.}~\bibnamefont {Simonovi\'c}},\ }\href {\doibase 10.1103/PhysRevD.102.063533} {\bibfield  {journal} {\bibinfo  {journal} {Phys. Rev. D}\ }\textbf {\bibinfo {volume} {102}},\ \bibinfo {pages} {063533} (\bibinfo {year} {2020})},\ \Eprint {http://arxiv.org/abs/2004.10607} {arXiv:2004.10607 [astro-ph.CO]} \BibitemShut {NoStop}%
\bibitem [{\citenamefont {Taule}\ and\ \citenamefont {Garny}(2023)}]{Taule:2023izt}%
  \BibitemOpen
  \bibfield  {author} {\bibinfo {author} {\bibfnamefont {P.}~\bibnamefont {Taule}}\ and\ \bibinfo {author} {\bibfnamefont {M.}~\bibnamefont {Garny}},\ }\href {\doibase 10.1088/1475-7516/2023/11/078} {\bibfield  {journal} {\bibinfo  {journal} {JCAP}\ }\textbf {\bibinfo {volume} {11}},\ \bibinfo {pages} {078} (\bibinfo {year} {2023})},\ \Eprint {http://arxiv.org/abs/2308.07379} {arXiv:2308.07379 [astro-ph.CO]} \BibitemShut {NoStop}%
\bibitem [{\citenamefont {Jackson}(1972)}]{Jackson:2008yv}%
  \BibitemOpen
  \bibfield  {author} {\bibinfo {author} {\bibfnamefont {J.~C.}\ \bibnamefont {Jackson}},\ }\href {\doibase 10.1093/mnras/156.1.1P} {\bibfield  {journal} {\bibinfo  {journal} {Mon. Not. Roy. Astron. Soc.}\ }\textbf {\bibinfo {volume} {156}},\ \bibinfo {pages} {1P} (\bibinfo {year} {1972})},\ \Eprint {http://arxiv.org/abs/0810.3908} {arXiv:0810.3908 [astro-ph]} \BibitemShut {NoStop}%
\bibitem [{\citenamefont {Perko}\ \emph {et~al.}(2016)\citenamefont {Perko}, \citenamefont {Senatore}, \citenamefont {Jennings},\ and\ \citenamefont {Wechsler}}]{Perko:2016puo}%
  \BibitemOpen
  \bibfield  {author} {\bibinfo {author} {\bibfnamefont {A.}~\bibnamefont {Perko}}, \bibinfo {author} {\bibfnamefont {L.}~\bibnamefont {Senatore}}, \bibinfo {author} {\bibfnamefont {E.}~\bibnamefont {Jennings}}, \ and\ \bibinfo {author} {\bibfnamefont {R.~H.}\ \bibnamefont {Wechsler}},\ }\href@noop {} {\  (\bibinfo {year} {2016})},\ \Eprint {http://arxiv.org/abs/1610.09321} {arXiv:1610.09321 [astro-ph.CO]} \BibitemShut {NoStop}%
\bibitem [{\citenamefont {Dawson}\ \emph {et~al.}(2013)\citenamefont {Dawson} \emph {et~al.}}]{BOSS:2012dmf}%
  \BibitemOpen
  \bibfield  {author} {\bibinfo {author} {\bibfnamefont {K.~S.}\ \bibnamefont {Dawson}} \emph {et~al.} (\bibinfo {collaboration} {BOSS}),\ }\href {\doibase 10.1088/0004-6256/145/1/10} {\bibfield  {journal} {\bibinfo  {journal} {Astron. J.}\ }\textbf {\bibinfo {volume} {145}},\ \bibinfo {pages} {10} (\bibinfo {year} {2013})},\ \Eprint {http://arxiv.org/abs/1208.0022} {arXiv:1208.0022 [astro-ph.CO]} \BibitemShut {NoStop}%
\bibitem [{\citenamefont {Ivanov}\ \emph {et~al.}(2022{\natexlab{b}})\citenamefont {Ivanov}, \citenamefont {Philcox}, \citenamefont {Simonovi\'c}, \citenamefont {Zaldarriaga}, \citenamefont {Nischimichi},\ and\ \citenamefont {Takada}}]{Ivanov:2021fbu}%
  \BibitemOpen
  \bibfield  {author} {\bibinfo {author} {\bibfnamefont {M.~M.}\ \bibnamefont {Ivanov}}, \bibinfo {author} {\bibfnamefont {O.~H.~E.}\ \bibnamefont {Philcox}}, \bibinfo {author} {\bibfnamefont {M.}~\bibnamefont {Simonovi\'c}}, \bibinfo {author} {\bibfnamefont {M.}~\bibnamefont {Zaldarriaga}}, \bibinfo {author} {\bibfnamefont {T.}~\bibnamefont {Nischimichi}}, \ and\ \bibinfo {author} {\bibfnamefont {M.}~\bibnamefont {Takada}},\ }\href {\doibase 10.1103/PhysRevD.105.043531} {\bibfield  {journal} {\bibinfo  {journal} {Phys. Rev. D}\ }\textbf {\bibinfo {volume} {105}},\ \bibinfo {pages} {043531} (\bibinfo {year} {2022}{\natexlab{b}})},\ \Eprint {http://arxiv.org/abs/2110.00006} {arXiv:2110.00006 [astro-ph.CO]} \BibitemShut {NoStop}%
\bibitem [{\citenamefont {Philcox}\ \emph {et~al.}(2020)\citenamefont {Philcox}, \citenamefont {Ivanov}, \citenamefont {Simonovi\'c},\ and\ \citenamefont {Zaldarriaga}}]{Philcox:2020vvt}%
  \BibitemOpen
  \bibfield  {author} {\bibinfo {author} {\bibfnamefont {O.~H.~E.}\ \bibnamefont {Philcox}}, \bibinfo {author} {\bibfnamefont {M.~M.}\ \bibnamefont {Ivanov}}, \bibinfo {author} {\bibfnamefont {M.}~\bibnamefont {Simonovi\'c}}, \ and\ \bibinfo {author} {\bibfnamefont {M.}~\bibnamefont {Zaldarriaga}},\ }\href {\doibase 10.1088/1475-7516/2020/05/032} {\bibfield  {journal} {\bibinfo  {journal} {JCAP}\ }\textbf {\bibinfo {volume} {05}},\ \bibinfo {pages} {032} (\bibinfo {year} {2020})},\ \Eprint {http://arxiv.org/abs/2002.04035} {arXiv:2002.04035 [astro-ph.CO]} \BibitemShut {NoStop}%
\bibitem [{\citenamefont {Chudaykin}\ \emph {et~al.}(2024)\citenamefont {Chudaykin}, \citenamefont {Ivanov},\ and\ \citenamefont {Nishimichi}}]{Chudaykin:2024wlw}%
  \BibitemOpen
  \bibfield  {author} {\bibinfo {author} {\bibfnamefont {A.}~\bibnamefont {Chudaykin}}, \bibinfo {author} {\bibfnamefont {M.~M.}\ \bibnamefont {Ivanov}}, \ and\ \bibinfo {author} {\bibfnamefont {T.}~\bibnamefont {Nishimichi}},\ }\href@noop {} {\  (\bibinfo {year} {2024})},\ \Eprint {http://arxiv.org/abs/2410.16358} {arXiv:2410.16358 [astro-ph.CO]} \BibitemShut {NoStop}%
\bibitem [{\citenamefont {Philcox}(2021{\natexlab{a}})}]{Philcox:2020vbm}%
  \BibitemOpen
  \bibfield  {author} {\bibinfo {author} {\bibfnamefont {O.~H.~E.}\ \bibnamefont {Philcox}},\ }\href {\doibase 10.1103/PhysRevD.103.103504} {\bibfield  {journal} {\bibinfo  {journal} {Phys. Rev. D}\ }\textbf {\bibinfo {volume} {103}},\ \bibinfo {pages} {103504} (\bibinfo {year} {2021}{\natexlab{a}})},\ \Eprint {http://arxiv.org/abs/2012.09389} {arXiv:2012.09389 [astro-ph.CO]} \BibitemShut {NoStop}%
\bibitem [{\citenamefont {Philcox}(2021{\natexlab{b}})}]{Philcox:2021ukg}%
  \BibitemOpen
  \bibfield  {author} {\bibinfo {author} {\bibfnamefont {O.~H.~E.}\ \bibnamefont {Philcox}},\ }\href {\doibase 10.1103/PhysRevD.104.123529} {\bibfield  {journal} {\bibinfo  {journal} {Phys. Rev. D}\ }\textbf {\bibinfo {volume} {104}},\ \bibinfo {pages} {123529} (\bibinfo {year} {2021}{\natexlab{b}})},\ \Eprint {http://arxiv.org/abs/2107.06287} {arXiv:2107.06287 [astro-ph.CO]} \BibitemShut {NoStop}%
\bibitem [{\citenamefont {Adame}\ \emph {et~al.}(2024{\natexlab{b}})\citenamefont {Adame} \emph {et~al.}}]{DESI:2024jis}%
  \BibitemOpen
  \bibfield  {author} {\bibinfo {author} {\bibfnamefont {A.~G.}\ \bibnamefont {Adame}} \emph {et~al.} (\bibinfo {collaboration} {DESI}),\ }\href@noop {} {\  (\bibinfo {year} {2024}{\natexlab{b}})},\ \Eprint {http://arxiv.org/abs/2411.12021} {arXiv:2411.12021 [astro-ph.CO]} \BibitemShut {NoStop}%
\bibitem [{\citenamefont {Aghanim}\ \emph {et~al.}(2020{\natexlab{b}})\citenamefont {Aghanim} \emph {et~al.}}]{Planck:2019nip}%
  \BibitemOpen
  \bibfield  {author} {\bibinfo {author} {\bibfnamefont {N.}~\bibnamefont {Aghanim}} \emph {et~al.} (\bibinfo {collaboration} {Planck}),\ }\href {\doibase 10.1051/0004-6361/201936386} {\bibfield  {journal} {\bibinfo  {journal} {Astron. Astrophys.}\ }\textbf {\bibinfo {volume} {641}},\ \bibinfo {pages} {A5} (\bibinfo {year} {2020}{\natexlab{b}})},\ \Eprint {http://arxiv.org/abs/1907.12875} {arXiv:1907.12875 [astro-ph.CO]} \BibitemShut {NoStop}%
\bibitem [{\citenamefont {Aghanim}\ \emph {et~al.}(2020{\natexlab{c}})\citenamefont {Aghanim} \emph {et~al.}}]{Planck:2018nkj}%
  \BibitemOpen
  \bibfield  {author} {\bibinfo {author} {\bibfnamefont {N.}~\bibnamefont {Aghanim}} \emph {et~al.} (\bibinfo {collaboration} {Planck}),\ }\href {\doibase 10.1051/0004-6361/201833880} {\bibfield  {journal} {\bibinfo  {journal} {Astron. Astrophys.}\ }\textbf {\bibinfo {volume} {641}},\ \bibinfo {pages} {A1} (\bibinfo {year} {2020}{\natexlab{c}})},\ \Eprint {http://arxiv.org/abs/1807.06205} {arXiv:1807.06205 [astro-ph.CO]} \BibitemShut {NoStop}%
\bibitem [{\citenamefont {Albrecht}\ \emph {et~al.}(2006)\citenamefont {Albrecht} \emph {et~al.}}]{Albrecht:2006um}%
  \BibitemOpen
  \bibfield  {author} {\bibinfo {author} {\bibfnamefont {A.}~\bibnamefont {Albrecht}} \emph {et~al.},\ }\href@noop {} {\  (\bibinfo {year} {2006})},\ \Eprint {http://arxiv.org/abs/astro-ph/0609591} {arXiv:astro-ph/0609591} \BibitemShut {NoStop}%
\bibitem [{\citenamefont {Tang}\ \emph {et~al.}(2025)\citenamefont {Tang}, \citenamefont {Brout}, \citenamefont {Karwal}, \citenamefont {Chang}, \citenamefont {Miranda},\ and\ \citenamefont {Vincenzi}}]{Tang:2024lmo}%
  \BibitemOpen
  \bibfield  {author} {\bibinfo {author} {\bibfnamefont {X.~T.}\ \bibnamefont {Tang}}, \bibinfo {author} {\bibfnamefont {D.}~\bibnamefont {Brout}}, \bibinfo {author} {\bibfnamefont {T.}~\bibnamefont {Karwal}}, \bibinfo {author} {\bibfnamefont {C.}~\bibnamefont {Chang}}, \bibinfo {author} {\bibfnamefont {V.}~\bibnamefont {Miranda}}, \ and\ \bibinfo {author} {\bibfnamefont {M.}~\bibnamefont {Vincenzi}},\ }\href {\doibase 10.3847/2041-8213/adc4da} {\bibfield  {journal} {\bibinfo  {journal} {Astrophys. J. Lett.}\ }\textbf {\bibinfo {volume} {983}},\ \bibinfo {pages} {L27} (\bibinfo {year} {2025})},\ \Eprint {http://arxiv.org/abs/2412.04430} {arXiv:2412.04430 [astro-ph.CO]} \BibitemShut {NoStop}%
\bibitem [{\citenamefont {Sailer}\ \emph {et~al.}(2025)\citenamefont {Sailer}, \citenamefont {Farren}, \citenamefont {Ferraro},\ and\ \citenamefont {White}}]{Sailer:2025lxj}%
  \BibitemOpen
  \bibfield  {author} {\bibinfo {author} {\bibfnamefont {N.}~\bibnamefont {Sailer}}, \bibinfo {author} {\bibfnamefont {G.~S.}\ \bibnamefont {Farren}}, \bibinfo {author} {\bibfnamefont {S.}~\bibnamefont {Ferraro}}, \ and\ \bibinfo {author} {\bibfnamefont {M.}~\bibnamefont {White}},\ }\href@noop {} {\  (\bibinfo {year} {2025})},\ \Eprint {http://arxiv.org/abs/2504.16932} {arXiv:2504.16932 [astro-ph.CO]} \BibitemShut {NoStop}%
\bibitem [{\citenamefont {Xu}\ \emph {et~al.}(2022)\citenamefont {Xu}, \citenamefont {Mu\~noz},\ and\ \citenamefont {Dvorkin}}]{Xu:2021rwg}%
  \BibitemOpen
  \bibfield  {author} {\bibinfo {author} {\bibfnamefont {W.~L.}\ \bibnamefont {Xu}}, \bibinfo {author} {\bibfnamefont {J.~B.}\ \bibnamefont {Mu\~noz}}, \ and\ \bibinfo {author} {\bibfnamefont {C.}~\bibnamefont {Dvorkin}},\ }\href {\doibase 10.1103/PhysRevD.105.095029} {\bibfield  {journal} {\bibinfo  {journal} {Phys. Rev. D}\ }\textbf {\bibinfo {volume} {105}},\ \bibinfo {pages} {095029} (\bibinfo {year} {2022})},\ \Eprint {http://arxiv.org/abs/2107.09664} {arXiv:2107.09664 [astro-ph.CO]} \BibitemShut {NoStop}%
\bibitem [{\citenamefont {Rogers}\ \emph {et~al.}(2023)\citenamefont {Rogers}, \citenamefont {Hlo\v{z}ek}, \citenamefont {Lagu\"e}, \citenamefont {Ivanov}, \citenamefont {Philcox}, \citenamefont {Cabass}, \citenamefont {Akitsu},\ and\ \citenamefont {Marsh}}]{Rogers:2023ezo}%
  \BibitemOpen
  \bibfield  {author} {\bibinfo {author} {\bibfnamefont {K.~K.}\ \bibnamefont {Rogers}}, \bibinfo {author} {\bibfnamefont {R.}~\bibnamefont {Hlo\v{z}ek}}, \bibinfo {author} {\bibfnamefont {A.}~\bibnamefont {Lagu\"e}}, \bibinfo {author} {\bibfnamefont {M.~M.}\ \bibnamefont {Ivanov}}, \bibinfo {author} {\bibfnamefont {O.~H.~E.}\ \bibnamefont {Philcox}}, \bibinfo {author} {\bibfnamefont {G.}~\bibnamefont {Cabass}}, \bibinfo {author} {\bibfnamefont {K.}~\bibnamefont {Akitsu}}, \ and\ \bibinfo {author} {\bibfnamefont {D.~J.~E.}\ \bibnamefont {Marsh}},\ }\href {\doibase 10.1088/1475-7516/2023/06/023} {\bibfield  {journal} {\bibinfo  {journal} {JCAP}\ }\textbf {\bibinfo {volume} {06}},\ \bibinfo {pages} {023} (\bibinfo {year} {2023})},\ \Eprint {http://arxiv.org/abs/2301.08361} {arXiv:2301.08361 [astro-ph.CO]} \BibitemShut {NoStop}%
\bibitem [{\citenamefont {He}\ \emph {et~al.}(2023)\citenamefont {He}, \citenamefont {Ivanov}, \citenamefont {An},\ and\ \citenamefont {Gluscevic}}]{He:2023dbn}%
  \BibitemOpen
  \bibfield  {author} {\bibinfo {author} {\bibfnamefont {A.}~\bibnamefont {He}}, \bibinfo {author} {\bibfnamefont {M.~M.}\ \bibnamefont {Ivanov}}, \bibinfo {author} {\bibfnamefont {R.}~\bibnamefont {An}}, \ and\ \bibinfo {author} {\bibfnamefont {V.}~\bibnamefont {Gluscevic}},\ }\href {\doibase 10.3847/2041-8213/acdb63} {\bibfield  {journal} {\bibinfo  {journal} {Astrophys. J. Lett.}\ }\textbf {\bibinfo {volume} {954}},\ \bibinfo {pages} {L8} (\bibinfo {year} {2023})},\ \Eprint {http://arxiv.org/abs/2301.08260} {arXiv:2301.08260 [astro-ph.CO]} \BibitemShut {NoStop}%
\bibitem [{\citenamefont {He}\ \emph {et~al.}(2024)\citenamefont {He}, \citenamefont {An}, \citenamefont {Ivanov},\ and\ \citenamefont {Gluscevic}}]{He:2023oke}%
  \BibitemOpen
  \bibfield  {author} {\bibinfo {author} {\bibfnamefont {A.}~\bibnamefont {He}}, \bibinfo {author} {\bibfnamefont {R.}~\bibnamefont {An}}, \bibinfo {author} {\bibfnamefont {M.~M.}\ \bibnamefont {Ivanov}}, \ and\ \bibinfo {author} {\bibfnamefont {V.}~\bibnamefont {Gluscevic}},\ }\href {\doibase 10.1103/PhysRevD.109.103527} {\bibfield  {journal} {\bibinfo  {journal} {Phys. Rev. D}\ }\textbf {\bibinfo {volume} {109}},\ \bibinfo {pages} {103527} (\bibinfo {year} {2024})},\ \Eprint {http://arxiv.org/abs/2309.03956} {arXiv:2309.03956 [astro-ph.CO]} \BibitemShut {NoStop}%
\bibitem [{\citenamefont {He}\ \emph {et~al.}(2025{\natexlab{a}})\citenamefont {He}, \citenamefont {Ivanov}, \citenamefont {An}, \citenamefont {Driskell},\ and\ \citenamefont {Gluscevic}}]{He:2025npy}%
  \BibitemOpen
  \bibfield  {author} {\bibinfo {author} {\bibfnamefont {A.}~\bibnamefont {He}}, \bibinfo {author} {\bibfnamefont {M.~M.}\ \bibnamefont {Ivanov}}, \bibinfo {author} {\bibfnamefont {R.}~\bibnamefont {An}}, \bibinfo {author} {\bibfnamefont {T.}~\bibnamefont {Driskell}}, \ and\ \bibinfo {author} {\bibfnamefont {V.}~\bibnamefont {Gluscevic}},\ }\href {\doibase 10.1088/1475-7516/2025/05/087} {\bibfield  {journal} {\bibinfo  {journal} {JCAP}\ }\textbf {\bibinfo {volume} {05}},\ \bibinfo {pages} {087} (\bibinfo {year} {2025}{\natexlab{a}})},\ \Eprint {http://arxiv.org/abs/2502.02636} {arXiv:2502.02636 [astro-ph.CO]} \BibitemShut {NoStop}%
\bibitem [{\citenamefont {Ivanov}\ \emph {et~al.}(2020{\natexlab{c}})\citenamefont {Ivanov}, \citenamefont {McDonough}, \citenamefont {Hill}, \citenamefont {Simonovi\'c}, \citenamefont {Toomey}, \citenamefont {Alexander},\ and\ \citenamefont {Zaldarriaga}}]{Ivanov:2020ril}%
  \BibitemOpen
  \bibfield  {author} {\bibinfo {author} {\bibfnamefont {M.~M.}\ \bibnamefont {Ivanov}}, \bibinfo {author} {\bibfnamefont {E.}~\bibnamefont {McDonough}}, \bibinfo {author} {\bibfnamefont {J.~C.}\ \bibnamefont {Hill}}, \bibinfo {author} {\bibfnamefont {M.}~\bibnamefont {Simonovi\'c}}, \bibinfo {author} {\bibfnamefont {M.~W.}\ \bibnamefont {Toomey}}, \bibinfo {author} {\bibfnamefont {S.}~\bibnamefont {Alexander}}, \ and\ \bibinfo {author} {\bibfnamefont {M.}~\bibnamefont {Zaldarriaga}},\ }\href {\doibase 10.1103/PhysRevD.102.103502} {\bibfield  {journal} {\bibinfo  {journal} {Phys. Rev. D}\ }\textbf {\bibinfo {volume} {102}},\ \bibinfo {pages} {103502} (\bibinfo {year} {2020}{\natexlab{c}})},\ \Eprint {http://arxiv.org/abs/2006.11235} {arXiv:2006.11235 [astro-ph.CO]} \BibitemShut {NoStop}%
\bibitem [{\citenamefont {McDonough}\ \emph {et~al.}(2024)\citenamefont {McDonough}, \citenamefont {Hill}, \citenamefont {Ivanov}, \citenamefont {La~Posta},\ and\ \citenamefont {Toomey}}]{McDonough:2023qcu}%
  \BibitemOpen
  \bibfield  {author} {\bibinfo {author} {\bibfnamefont {E.}~\bibnamefont {McDonough}}, \bibinfo {author} {\bibfnamefont {J.~C.}\ \bibnamefont {Hill}}, \bibinfo {author} {\bibfnamefont {M.~M.}\ \bibnamefont {Ivanov}}, \bibinfo {author} {\bibfnamefont {A.}~\bibnamefont {La~Posta}}, \ and\ \bibinfo {author} {\bibfnamefont {M.~W.}\ \bibnamefont {Toomey}},\ }\href {\doibase 10.1142/S0218271824300039} {\bibfield  {journal} {\bibinfo  {journal} {Int. J. Mod. Phys. D}\ }\textbf {\bibinfo {volume} {33}},\ \bibinfo {pages} {2430003} (\bibinfo {year} {2024})},\ \Eprint {http://arxiv.org/abs/2310.19899} {arXiv:2310.19899 [astro-ph.CO]} \BibitemShut {NoStop}%
\bibitem [{\citenamefont {Toomey}\ \emph {et~al.}(2024)\citenamefont {Toomey}, \citenamefont {Ivanov},\ and\ \citenamefont {McDonough}}]{Toomey:2024ita}%
  \BibitemOpen
  \bibfield  {author} {\bibinfo {author} {\bibfnamefont {M.~W.}\ \bibnamefont {Toomey}}, \bibinfo {author} {\bibfnamefont {M.~M.}\ \bibnamefont {Ivanov}}, \ and\ \bibinfo {author} {\bibfnamefont {E.}~\bibnamefont {McDonough}},\ }\href@noop {} {\  (\bibinfo {year} {2024})},\ \Eprint {http://arxiv.org/abs/2409.09029} {arXiv:2409.09029 [astro-ph.CO]} \BibitemShut {NoStop}%
\bibitem [{\citenamefont {Ivanov}(2024)}]{Ivanov:2023yla}%
  \BibitemOpen
  \bibfield  {author} {\bibinfo {author} {\bibfnamefont {M.~M.}\ \bibnamefont {Ivanov}},\ }\href {\doibase 10.1103/PhysRevD.109.023507} {\bibfield  {journal} {\bibinfo  {journal} {Phys. Rev. D}\ }\textbf {\bibinfo {volume} {109}},\ \bibinfo {pages} {023507} (\bibinfo {year} {2024})},\ \Eprint {http://arxiv.org/abs/2309.10133} {arXiv:2309.10133 [astro-ph.CO]} \BibitemShut {NoStop}%
\bibitem [{\citenamefont {Ivanov}\ \emph {et~al.}(2024{\natexlab{c}})\citenamefont {Ivanov}, \citenamefont {Toomey},\ and\ \citenamefont {Kara\c{c}ayl\i{}}}]{Ivanov:2024jtl}%
  \BibitemOpen
  \bibfield  {author} {\bibinfo {author} {\bibfnamefont {M.~M.}\ \bibnamefont {Ivanov}}, \bibinfo {author} {\bibfnamefont {M.~W.}\ \bibnamefont {Toomey}}, \ and\ \bibinfo {author} {\bibfnamefont {N.~G.}\ \bibnamefont {Kara\c{c}ayl\i{}}},\ }\href@noop {} {\  (\bibinfo {year} {2024}{\natexlab{c}})},\ \Eprint {http://arxiv.org/abs/2405.13208} {arXiv:2405.13208 [astro-ph.CO]} \BibitemShut {NoStop}%
\bibitem [{\citenamefont {de~Belsunce}\ \emph {et~al.}(2025)\citenamefont {de~Belsunce}, \citenamefont {Chen}, \citenamefont {Ivanov}, \citenamefont {Ravoux}, \citenamefont {Chabanier}, \citenamefont {Sexton},\ and\ \citenamefont {Lukic}}]{deBelsunce:2024rvv}%
  \BibitemOpen
  \bibfield  {author} {\bibinfo {author} {\bibfnamefont {R.}~\bibnamefont {de~Belsunce}}, \bibinfo {author} {\bibfnamefont {S.-F.}\ \bibnamefont {Chen}}, \bibinfo {author} {\bibfnamefont {M.~M.}\ \bibnamefont {Ivanov}}, \bibinfo {author} {\bibfnamefont {C.}~\bibnamefont {Ravoux}}, \bibinfo {author} {\bibfnamefont {S.}~\bibnamefont {Chabanier}}, \bibinfo {author} {\bibfnamefont {J.}~\bibnamefont {Sexton}}, \ and\ \bibinfo {author} {\bibfnamefont {Z.}~\bibnamefont {Lukic}},\ }\href {\doibase 10.1103/PhysRevD.111.063524} {\bibfield  {journal} {\bibinfo  {journal} {Phys. Rev. D}\ }\textbf {\bibinfo {volume} {111}},\ \bibinfo {pages} {063524} (\bibinfo {year} {2025})},\ \Eprint {http://arxiv.org/abs/2412.06892} {arXiv:2412.06892 [astro-ph.CO]} \BibitemShut {NoStop}%
\bibitem [{\citenamefont {Chudaykin}\ and\ \citenamefont {Ivanov}(2025)}]{Chudaykin:2025gsh}%
  \BibitemOpen
  \bibfield  {author} {\bibinfo {author} {\bibfnamefont {A.}~\bibnamefont {Chudaykin}}\ and\ \bibinfo {author} {\bibfnamefont {M.~M.}\ \bibnamefont {Ivanov}},\ }\href {\doibase 10.1103/PhysRevD.111.083515} {\bibfield  {journal} {\bibinfo  {journal} {Phys. Rev. D}\ }\textbf {\bibinfo {volume} {111}},\ \bibinfo {pages} {083515} (\bibinfo {year} {2025})},\ \Eprint {http://arxiv.org/abs/2501.04770} {arXiv:2501.04770 [astro-ph.CO]} \BibitemShut {NoStop}%
\bibitem [{\citenamefont {He}\ \emph {et~al.}(2025{\natexlab{b}})\citenamefont {He}, \citenamefont {Ivanov}, \citenamefont {Bird}, \citenamefont {An},\ and\ \citenamefont {Gluscevic}}]{He:2025jwp}%
  \BibitemOpen
  \bibfield  {author} {\bibinfo {author} {\bibfnamefont {A.}~\bibnamefont {He}}, \bibinfo {author} {\bibfnamefont {M.~M.}\ \bibnamefont {Ivanov}}, \bibinfo {author} {\bibfnamefont {S.}~\bibnamefont {Bird}}, \bibinfo {author} {\bibfnamefont {R.}~\bibnamefont {An}}, \ and\ \bibinfo {author} {\bibfnamefont {V.}~\bibnamefont {Gluscevic}},\ }\href@noop {} {\  (\bibinfo {year} {2025}{\natexlab{b}})},\ \Eprint {http://arxiv.org/abs/2503.15592} {arXiv:2503.15592 [astro-ph.CO]} \BibitemShut {NoStop}%
\bibitem [{\citenamefont {Hadzhiyska}\ \emph {et~al.}(2025)\citenamefont {Hadzhiyska}, \citenamefont {de~Belsunce}, \citenamefont {Cuceu}, \citenamefont {Guy}, \citenamefont {Ivanov}, \citenamefont {Coquinot},\ and\ \citenamefont {Font-Ribera}}]{Hadzhiyska:2025cvk}%
  \BibitemOpen
  \bibfield  {author} {\bibinfo {author} {\bibfnamefont {B.}~\bibnamefont {Hadzhiyska}}, \bibinfo {author} {\bibfnamefont {R.}~\bibnamefont {de~Belsunce}}, \bibinfo {author} {\bibfnamefont {A.}~\bibnamefont {Cuceu}}, \bibinfo {author} {\bibfnamefont {J.}~\bibnamefont {Guy}}, \bibinfo {author} {\bibfnamefont {M.~M.}\ \bibnamefont {Ivanov}}, \bibinfo {author} {\bibfnamefont {H.}~\bibnamefont {Coquinot}}, \ and\ \bibinfo {author} {\bibfnamefont {A.}~\bibnamefont {Font-Ribera}},\ }\href@noop {} {\  (\bibinfo {year} {2025})},\ \Eprint {http://arxiv.org/abs/2503.13442} {arXiv:2503.13442 [astro-ph.CO]} \BibitemShut {NoStop}%
\end{thebibliography}%
\end{document}